\documentclass[12pt]{article}

\usepackage[left=1in,top=1in,right=1in,bottom=1in,head=.1in,nofoot]{geometry}
\usepackage{setspace,url,bm,amsmath} 
\usepackage{titlesec} 
\titlelabel{\thetitle.\quad} 
\usepackage[margin=20pt]{subcaption}
\usepackage{amsmath,amsfonts,amsthm,amssymb}
\usepackage{graphicx}
\usepackage[colorlinks=true,linkcolor=black,citecolor=blue,urlcolor=blue]{hyperref}
\usepackage[table]{xcolor}
\usepackage{multirow}
\usepackage{array}
\usepackage{booktabs}
\usepackage{threeparttable}
\usepackage{comment}
\usepackage{float}
\usepackage{natbib}
\usepackage{indentfirst}

\newtheorem*{theorem*}{Theorem}
\newtheorem{theorem}{Theorem}

\newtheorem*{corollary*}{Corollary}

\usepackage[ruled,vlined]{algorithm2e}
\usepackage{booktabs,multirow,caption,threeparttable}

\usepackage{enumitem}
\setenumerate[1]{itemsep=0pt,partopsep=0pt,parsep=\parskip,topsep=5pt}
\setitemize[1]{itemsep=0pt,partopsep=0pt,parsep=\parskip,topsep=5pt}

\def\htau{\theta}
\def\tl{\theta_l}
\def\Wrep{W^b}
\def\cardW{|\mathcal{W}_{a}|}
\def\WMC{\mathcal{W}_{a}^\mathrm{MC}}

\def\T{{ \mathrm{\scriptscriptstyle T} }}
\def\pr{P}
\def\Var{\mathrm{var}}
\def\Cov{\mathrm{cov}}

\def\imp{\mathrm{imp}}
\def\obs{\mathrm{obs}}

\def\cre{\mathrm{CR}}
\def\rr{\mathrm{RR}}
\def\psrr{\mathrm{PSRR}}

\def\seqpsrr{\mathrm{SeqPSRR}}

\def\R{\mathcal{R}}
\def\r{r}
\def\pcr{p_{\cre}}
\def\prr{p_r}

\begin{document}
\begin{singlespace}

\title{\bf Pair-switching rerandomization}
\author{
{
Ke Zhu and Hanzhong Liu\thanks{\small{Corresponding author. Email: \texttt{lhz2016@tsinghua.edu.cn}}}
}
\\ \\
{Center for Statistical Science, Department of Industrial Engineering,}  \\ {Tsinghua University, Beijing, China}
}

\date{}
\maketitle
\end{singlespace}

\thispagestyle{empty}
\vskip -8mm

\begin{singlespace}
\begin{abstract}
Rerandomization discards assignments with covariates unbalanced in the treatment and control groups to improve estimation and inference efficiency. However, the acceptance-rejection sampling method used in rerandomization is computationally inefficient. As a result, it is time-consuming for rerandomization to draw numerous independent assignments, which are necessary for performing Fisher randomization tests and constructing randomization-based confidence intervals. To address this problem, we propose a pair-switching rerandomization method to draw balanced assignments efficiently. We obtain the unbiasedness and variance reduction of the difference-in-means estimator and show that the Fisher randomization tests are valid under pair-switching rerandomization. Moreover, we propose an exact approach to invert Fisher randomization tests to confidence intervals, which is faster than the existing methods. In addition, our method is applicable to both non-sequentially and sequentially randomized experiments. We conduct comprehensive simulation studies to compare the finite-sample performance of the proposed method with that of classical rerandomization. Simulation results indicate that pair-switching rerandomization leads to comparable power of Fisher randomization tests and is 3--23 times faster than classical rerandomization. Finally, we apply the pair-switching rerandomization method to analyze two clinical trial datasets, both of which demonstrate the advantages of our method.

\vspace{12pt}
\end{abstract}

\end{singlespace}

\newpage

\clearpage
\setcounter{page}{1}

\allowdisplaybreaks

\baselineskip=24pt

\section{Introduction}\label{sec:intro}

Randomized experiments are the gold standard for drawing causal inference because all observed and unobserved confounders are balanced on average by randomization. However, even in completely randomized experiments, there is a high chance of producing an unbalanced assignment \citep{rosen2008, rubin2008comment, xu2010}, and the probability of producing such an unbalanced assignment increases with the number of covariates \citep{morgan2012,krieger2019bka}. 
Some researchers advocate balancing covariates in the design stage to improve the credibility of a study and increase the precision of estimation and power of tests \citep[e.g.,][]{student1938comparison,greevy2004,kallus2018,harshaw2020}. \citet{fisher1926} proposed to use stratification to balance a few  categorical covariates that are most relevant to the outcomes. When there are many important covariates, a large class of covariate-adaptive or covariate-adjusted response-adaptive randomization methods has been proposed to achieve covariate balance across treatment groups by sequentially modifying the probabilities of assignments \citep{taves1974min, pocock1975seq, rosen2008, hu2012asy, hu2014adap}. If the covariates of all units can be collected before the physical implementation of experiments, such as in many phase 1 clinical trials \citep{senn2013se} and randomized controlled trials in economics \citep{athey2017econ, banerjee2020}, units can be repeatedly assigned to treatment groups until all categorical and continuous covariates are satisfactorily balanced, which is often called rerandomization. For example, \citet{maclure2006measuring} used rerandomization in a clinical trial to assess the influence of physician education tools (PET) on the prescription quality for general practitioners. According to a survey \citep{bruhn2009pursuit}, rerandomization is commonly used in the design stage of policy evaluation to balance key covariates but is less commonly documented \citep{heckman2021}.

Although balanced designs can improve efficiency, scholars advocate against deterministic designs that sacrifice randomness for the sake of better balance \citep{efron1971forcing,kapelner2021harmonizing,johansson2021}. If we force the exact balance within each (small) block in a small trial, knowing past allocations allows for the accurate prediction of most future allocations in the same block, which may introduce selection bias into the trial \citep[][Section 5]{rosenberger2015rand}.
Moreover, deterministic designs do not enjoy the merits of randomization as a reasoned basis for inference \citep[][Section 6]{rosenberger2015rand}. For further critiques of deterministic designs, we refer readers to \cite{rosen2008} and \cite{senn2013se}.

Unlike deterministic designs, rerandomization maintains randomization and can be seen as ``a harmony of optimal deterministic design and completely randomized design'' \citep{kapelner2021harmonizing}.
\citet{morgan2012} formally investigated the theoretical properties of rerandomization using the Mahalanobis distance. This rerandomization procedure only accepts those assignments with the Mahalanobis distance of the covariate means between the treatment and control groups less than or equal to a prespecified threshold. They derived the validity and efficiency gains of rerandomization under the conditions of additive treatment effects and equal treatment and control group sizes. Moreover, they proposed the use of Fisher randomization tests \citep{fisher1935} to take the rerandomization into account in the analysis stage and demonstrated its validity. Under more general conditions, \cite{li2018} derived the asymptotic distribution of the difference-in-means estimator under rerandomization and proposed an asymptotically valid inference method without imposing any parametric modelling assumptions on the potential outcomes and covariates.

When the sample size was small or moderate, the simulation studies in \citet{johansson2021} showed that under rerandomization, the asymptotic test proposed by \citet{li2018} failed to control the type one error. In contrast, the Fisher randomization tests performed well. In general, many scholars advocate Fisher randomization tests as a credible and flexible inference approach over asymptotic inference based on extensive empirical studies \citep{keele2015statistics,young2019, proschan2019re,bind2020} and theoretical analyses \citep{wu2020,zhao2021,cohen2020,luo2021,caughey2021rand,branson2021rand}.

However, it is computationally challenging to perform Fisher randomization tests under rerandomization \citep{luo2021}. In fact, Fisher randomization tests are computationally intensive, even under complete randomization \citep{chung2018rapid,bind2020}. Under rerandomization, generating only one well-balanced assignment often requires drawing thousands of assignments, not to mention, sampling numerous well-balanced assignments to conduct Fisher randomization tests. The computational issue is even more severe when we construct randomization-based confidence intervals by inverting a series of Fisher randomization tests \citep[][Section 5.7]{imbens2015}.

Modern sampling techniques, such as Markov Chain Monte Carlo, help improve the sampling efficiency \citep{liu2008monte, givens2013computational}. Motivated by the Metropolis--Hasting algorithm \citep{metropolis1953equation, hastings1970monte}, our first contribution is to propose a pair-switching rerandomization (PSRR) method, which can save a huge amount of computational cost of classical rerandomization. Our main idea is to sample an acceptable (well-balanced) assignment along the path of gradual improvement in covariate balance. Specifically, we start with a completely randomized assignment. If it is not acceptable, then we try to move, or ``rerandomize,'' to a more balanced assignment by ``pair-switching''; that is, switching the treatment status of two randomly selected units--one is in the treatment group, and the other is in the control group. To prevent trapping in a local optimum, we allow moving to less-balanced assignments with specific probabilities. We switch until we find an acceptable assignment. Our simulations and real data analysis illustrate that PSRR is approximately 22--23 times faster than classical rerandomization to achieve comparable powers of Fisher randomization tests. Furthermore, we obtain the unbiasedness and a lower bound on the variance reduction of the difference-in-means estimator under PSRR.

Most relevant to our study, \citet{krieger2019bka} proposed a greedy pair-switching (GPS) algorithm to improve the balance performance of classical rerandomization. Rather than sampling an assignment under which covariate balance is achieved (the Mahalanobis distance is less than a prespecified threshold), GPS tries to find an assignment with locally optimal balance by greedy pair-switching. In particular, in the case of dividing $n$ units into equal-sized treatment and control groups, GPS starts with a random assignment and then moves to the \textit{most balanced} assignment of all $(n/2)^2$ tentative pair-switching assignments until no pair-switching can further reduce the imbalance. In contrast, PSRR allows us to move to \textit{less-balanced assignments with specific probabilities}. Thus, PSRR is a non-greedy heuristic, which shares elements with other heuristics, such as the simulated annealing algorithm \citep[][Section 3.3]{givens2013computational} and epsilon-greedy algorithm \citep[][Section 2.2]{sutton2018re}. By sacrificing short-term benefits, the non-greedy heuristic explores the space more fully and prevents trapping in the local optimum. In our simulation, PSRR only tries 2--5 times to move to a new candidate assignment and is approximately 4--110 times faster than GPS to achieve comparable powers of Fisher randomization tests.

Our second contribution is to propose an exact approach to construct randomization-based confidence intervals by inverting Fisher randomization tests. Existing methods determine the endpoints of the intervals using numerical approximations \citep{garthwaite1996,wang2020ci,luo2021}. In contrast, our approach finds the endpoints by solving a series of linear equations, whose computational cost is almost the same as that of a single Fisher randomization test.

In many clinical trials and A/B testing applications \citep{kapelner2014matching,qin2016pairwise,bertsimas2019,bhat2020}, the units enroll the experiment sequentially. The experimenter might be unable to wait to conduct the experiment until all the experimental units arrive. This motivates researchers to consider the sequential assignment of treatment status one by one (fully sequential experiments) or group by group (group sequential experiments). To improve the estimation and inference efficiency in group sequential experiments, \citet{zhou2018} proposed a sequential rerandomization method. However, sequential rerandomization is plagued by low computational efficiency. Our third contribution is the generalization of PSRR in group sequential experiments. Sequential pair-switching rerandomization is 3--7 times faster than sequential rerandomization with comparable powers.

The remainder of this paper is organized as follows. In Section \ref{sec:framework}, we introduce the potential outcomes framework, rerandomization, Fisher randomization tests, and randomization-based confidence intervals. In Section \ref{sec:psrr}, we propose a PSRR method and study its theoretical properties. 
We extend our method to sequentially randomized experiments in Section \ref{sec:seq}.
In Section \ref{sec:sim}, we conduct simulation studies to compare the performance of PSRR with that of existing methods.
In Section \ref{sec:real}, we illustrate our method using two clinical trial datasets. 
We conclude the paper with a discussion of future work in Section \ref{sec:dis}. 
All proofs and additional simulation results are provided in Web Appendices.

\section{Randomized experiments and rerandomization}
\label{sec:framework}

\subsection{Framework and notation}

To ground our discussion, we introduce a phase 1 clinical trial assessing the interactions between oral reserpine and intravenous methamphetamine. The goal of this study was to evaluate the safety of reserpine for the withdrawal of methamphetamine dependence. After all participants were recruited, 20 were randomly assigned to the treatment group, which received reserpine plus methamphetamine, and the remaining 10 participants received placebo plus methamphetamine (control group). The outcomes of interest were a series of pharmacological responses, including post-treatment heart rate, which we use for illustration purposes. 
Summary statistics of the dataset are shown in Table C7 in Web Appendix C.
The post-treatment heart rate of the treatment group was significantly lower than that of the control group, which is contrary to the prediction of the protocol. Moreover, some baseline covariates were not well-balanced in the sense that the standardized differences were outside the range $[-0.1,0.1]$ \citep{austin2009balance}. In particular, the pre-treatment heart rate of the treatment group was lower than that of the control group. Balancing these covariates may be helpful in improving the interpretability of this study and the estimation efficiency of the treatment effect.

In this study, we adopt the Neyman--Rubin potential outcome framework to define the treatment effect \citep{neyman1990, rubin1974}.
For the $i$th participant in the trial ($i=1,\dots,n$), we suppose that she/he has two potential outcomes, $Y_i(1)$ and $Y_i(0)$, which represent her/his post-treatment heart rates receiving reserpine plus methamphetamine and placebo plus methamphetamine, respectively. The unit-level treatment effect is defined as $\tau_i = Y_i(1) - Y_i(0)$. Since each participant can only receive reserpine plus methamphetamine or placebo plus methamphetamine, but not both, we can only observe one of the potential outcomes. Thus, the unit-level treatment effect is not identifiable without other modelling assumptions.
Fortunately, under the stable unit treatment value assumption (SUTVA) \citep{rubin1980}, which states that the potential outcomes of one unit are unaffected by the treatment status of other units and that there is only one version of each treatment status, we can estimate the average treatment effect, which is defined as $\tau= n^{-1} \sum_{i=1}^n(Y_i(1)-Y_i(0))$.
For each participant $i$, let us denote $W_i$ as her/his treatment status, $W_i=1$ if he/she received reserpine plus methamphetamine and $W_i=0$ otherwise. We denote the number of units in the treatment group as $n_t$, and the number of units in the control group as $n_c$. Let $W=(W_1,\dots,W_n)^\T$ be the vector of treatment assignment. In completely randomized experiments, the probability that $W$ takes a particular value $w=(w_1,\dots,w_n)$ with $w_i \in \{0,1\}$ is $\pr( W = w ) = n_t! n_c!/n!$, $\sum_{i=1}^{n}w_i = n_t$. Under SUTVA, the observed outcome of unit $i$ is $Y_i = W_iY_i(1) + (1-W_i)Y_i(0)$. For each participant $i$, we also observe a $p$-dimensional vector of baseline covariates $X_i=(X_{i1},\ldots,X_{ip})^\T$, such as age, sex, and pre-treatment heart rate. The covariate matrix is denoted by $X=(X_{1},\ldots,X_{n})^\T$. These covariates can be either continuous or categorical. Let $\overline{X}= n^{-1}\sum_{i=1}^{n} X_i$. Our goal is to infer $\tau$ using observed data $\{Y_i, W_i, X_i\}_{i=1}^{n}$.

\subsection{Rerandomization}
\label{sec:review}

\citet{morgan2012} suggested using the Mahalanobis distance of the covariate means in the treatment and control groups to measure the covariate balance.
For a given assignment $W$, the Mahalanobis distance is defined as
$$
\begin{aligned}
M(W)&\equiv\left(\overline X_t-\overline X_c\right)^{\T}\left[\Cov\left(\overline X_t-\overline X_c\right)\right]^{-1}\left(\overline X_t-\overline X_c\right) \\ 
&=n_t\left(1-n_t/n\right)\left(\overline X_t-\overline X_c\right)^\T S_{XX}^{-1}\left(\overline X_t-\overline X_c\right),
\end{aligned}
$$
where $\overline X_t=\sum_{i:W_i=1}X_{i}/n_t$ and $\overline X_c=\sum_{i:W_i=0}X_{i}/n_c$ are the mean vectors of the covariates in the treatment and control groups, respectively, and $S_{XX} = (n-1)^{-1} \sum_{i=1}^n (X_i - \overline{X})  (X_i - \overline{X})^\T $ is the covariance matrix of the covariates. The treatment assignment is acceptable if $M(W)\leq a$, where $a>0$ is a prespecified threshold. The whole procedure of rerandomization is as follows:
\begin{enumerate}
	\item Collect covariates data and specify the balance criterion as $M(W) \leq a $.
    \item Randomly generate a candidate assignment with $n_t$ units in the treatment group and $n_c$ units in the control group.
    \item Check whether the candidate assignment is acceptable by the balance criterion. If acceptable, proceed to Step 4. Otherwise, return to Step 2.
    \item Conduct the experiment using the acceptable assignment.
\end{enumerate}
We denote the set of acceptable assignments by $\mathcal{W}_a = \{W :  M(W) \leq a, \ \sum_{i=1}^n W_i=n_t\}$. When $a=\infty$, rerandomization is equivalent to complete randomization and $\mathcal{W}_{\infty} = \{W :  \sum_{i=1}^n W_i=n_t\}$.
The above procedure can be regarded as an acceptance-rejection sampling method: samples from $\mathcal{W}_{\infty}$ are accepted if they fall into $\mathcal{W}_{a}\subset\mathcal{W}_{\infty}$, and rejected otherwise.
\citet{morgan2012} proved that the asymptotic acceptance probability of the assignment is $p_a=\pr(\chi^2_p<a)$.
In practice, \citet{li2018} recommended that $p_a=0.001$, which means that we need to draw approximately $1/p_a=1000$ treatment assignments in $\mathcal{W}_{\infty}$ to find an acceptable assignment in $\mathcal{W}_{a}$.

\subsection{Fisher randomization tests}
\label{sec:frt}

As discussed in Section \ref{sec:intro}, after conducting the experiment using the assignment generated by rerandomization, we can use Fisher randomization tests to test the sharp null hypothesis, $H_0:Y_i(1)-Y_i(0)=0,\, i = 1, \ldots, n$. Under $H_0$, we can impute all potential outcomes by using the observed outcomes, $\widetilde Y_i(1)= \widetilde Y_i(0) = Y_i$.
We first select a test statistic to test $H_0$. A widely used test statistic is the difference-in-means estimator, $\widehat\tau(W)=\sum_{i:W_i=1} Y_i(1)/n_t - \sum_{i:W_i=0} Y_i(0)/n_c$.
Under $H_0$, the exact distribution of $\widehat\tau(W)$ is known; thus, we can calculate the exact $p$-value. However, the computation of the exact $p$-value is intensive when the cardinality of set $\mathcal{W}_a$, denoted by $\cardW$, is too large. In practice, we often use the Monte Carlo method to approximate the exact $p$-value. More specifically, we independently sample $B$ assignments $\{W^{b}, b=1,\dots,B\}$ following the rerandomization procedure and then calculate the corresponding values of the test statistic $\{\widehat\tau(W^{b}), b=1,\dots,B\}$ under $H_0$. For the two-sided alternative $H_1:Y_i(1)-Y_i(0)\neq 0$, larger values of $|\widehat\tau(W^{b})|$ indicate a departure from $H_0$ in favor of $H_1$. Thus, the approximated $p$-value is defined as the proportion of $|\widehat\tau(W^{b})|$ values larger than or equal to the observed $|\widehat\tau(W^\obs)|$; that is, $\widehat{p}= B^{-1}
\sum_{b=1}^B I_{\{|\widehat\tau(W^{b})|\geq|\widehat\tau(W^\obs)|\}}$, where $I_{ \{ \cdot \} }$ is an indicator function and $W^\obs$ is the observed assignment.
To obtain a good approximation of the $p$-value, we need $B$ to be sufficiently large, for example, $B=1000$.
When $p_a=0.001$, we need to draw, on average, $B/p_a = 10^6$ treatment assignments from $\mathcal{W}_{\infty}$ to obtain $B=1000$ acceptable assignments and then perform Fisher randomization tests with an accuracy of the $p$-value of approximately $1/B=10^{-3}$.

\subsection{Randomization-based confidence intervals}

Because of the dual relationship between hypothesis testing and interval estimation, we can construct randomization-based confidence intervals by inverting Fisher randomization tests \citep{lehmann1963nonparametric}. 
We first consider the construction of a lower confidence bound with confidence level $1-\alpha$, $0 < \alpha < 1$. For this purpose, we need to test $H_0^\htau:Y_i(1)-Y_i(0)=\htau,\, i = 1, \ldots, n$, versus $H_1^\htau:Y_i(1)-Y_i(0)>\htau,\, i = 1, \ldots, n$, where $\htau\in\mathbb{R}$ is a hypothetical average treatment effect. Under $H_0^\htau$, we impute the unobserved outcomes and denote the observed and imputed outcomes as $\mathbf{Y}_\htau^{\imp}=\{\widetilde{Y}_i(1), \widetilde{Y}_i(0),i=1,\ldots,n\}$. For any hypothetical $\Wrep\in\mathcal{W}_a$, the corresponding value of the test statistic is denoted by $\widehat{\tau}(\mathbf{Y}_\htau^{\imp},\Wrep)$. Then, the exact $p$-value function is defined as $p(\htau)= \cardW^{-1}\sum_{\Wrep\in\mathcal{W}_a} I_{\{\widehat{\tau}(\mathbf{Y}_\htau^{\imp},\Wrep)\geq\widehat{\tau}(\mathbf{Y}_\htau^{\imp},W^\obs)\}}$. Letting $\tl\equiv\sup\left\{\htau: p(\htau)\leq \alpha\right\}$,  \citet[][Proposition 2]{luo2021} showed that $\tl$ is a lower confidence bound; that is, $[\tl,\infty)$ covers the true average treatment effect $\tau$ with a probability of at least $1-\alpha$. Several approaches have been proposed to approximate $\tl$, such as the grid search method \citep{rosenbaum2002obs,imbens2015}, Robbins--Monro algorithm \citep{garthwaite1996,wang2020ci}, and bisection method \citep{wang2020ci,luo2021}. In this study, we propose an exact approach to obtain the lower confidence bound, which saves significant computational costs.

The existing methods only use the monotonicity of $p(\htau)$. However, for a test statistic that is non-decreasing with respect to $\htau$, such as the difference-in-means estimator, $p(\htau)$ is also a right-continuous step function with finite jump points. If we can exactly \textit{determine all these jump points}, we can recover $p(\htau)$ and obtain the exact lower confidence bound $\tl$. The observed value of the test statistic $\widehat{\tau}^\obs\equiv\widehat{\tau}(\mathbf{Y}_\htau^{\imp},W^\obs)$ \textit{does not change} with $\htau$. Moreover, for any hypothetical assignment $\Wrep\in\mathcal{W}_a$ and $\Wrep\neq W^\obs$, the value of the test statistic $\widehat{\tau}^b(\htau)\equiv\widehat{\tau}(\mathbf{Y}_\htau^{\imp},W^b)$ \textit{increases} with $\htau$. As $p(\htau)$ is the proportion of $\widehat{\tau}^b(\htau)$ greater than or equal to $\widehat{\tau}^\obs$, $p(\htau)$ jumps \textit{only if} one of the $\widehat{\tau}^b(\htau)$'s that is smaller than $\widehat{\tau}^\obs$ becomes equal to $\widehat{\tau}^\obs$. Therefore, the jump point of $p(\htau)$ is $\htau$ such that $\widehat{\tau}^b(\htau)$ is equal to $\widehat{\tau}^\obs$.

\begin{theorem}
\label{thm:ci}
For $\Wrep\in\mathcal{W}_a$, $\Wrep\neq W^\obs$, the solution of $\widehat{\tau}(\mathbf{Y}_\htau^{\imp},\Wrep)=\widehat{\tau}(\mathbf{Y}_\htau^{\imp},W^\obs)$ is:
$$
\htau_b=\frac{\sum_{W_i^\obs=1,\Wrep_i=0}Y_i(1)- \sum_{W_i^\obs=0,\Wrep_i=1}Y_i(0)} {\sum_{i=1}^n I_{\{W_i^\obs=1,\Wrep_i=0\}}}.
$$
For $\Wrep = W^\obs$, let $\htau_b = - \infty$. Let $(\htau_{(1)},\ldots,\htau_{(\cardW)})$ be the increasingly ordered value of $\htau_b$. Subsequently, the one-sided $1- \alpha$ lower confidence bound $\tl$ is $\htau_{(\lfloor \alpha \cardW \rfloor +1)}$.
\end{theorem}

When $\cardW$ is too large, we have to use the Monte Carlo approximation of the exact $p$-value function. Specifically, we independently sample $B$ assignments from $\mathcal{W}_a$ and approximate the exact $p$-value function by $\widehat{p}(\htau)= B^{-1}\sum_{W^b\in\WMC} I_{\{\widehat{\tau}(\mathbf{Y}_\htau^{\imp},\Wrep)\geq\widehat{\tau}(\mathbf{Y}_\htau^{\imp},W^\obs)\}}$, where $\WMC$ is the set of assignments generated by the Monte Carlo approximation. Then, we can approximate $\tl$ using $\widehat{\tl}\equiv\sup\left\{\htau: \widehat{p}(\htau)\leq \alpha\right\}$, which is obtained by replacing $\mathcal{W}_a$ in Theorem \ref{thm:ci} with $\WMC$.

Similarly, we can construct the one-sided $1-\alpha$ upper confidence bound. The endpoints of the $1-\alpha$ two-sided confidence interval are the one-sided $1 - \alpha/2$ lower and upper confidence bounds. The exact approach is not only useful for PSRR, but also applicable to other experimental designs.

\section{Pair-switching rerandomization}
\label{sec:psrr}

Our main goal is to sample treatment assignment $W$ from $\mathcal{W}_a$ more efficiently.
To this end, we start with a random assignment and move towards an acceptable assignment through a random walk chain, $(W^{(0)},\ldots,W^{(T)})$.
Specifically, when $t=0$, we sample a completely randomized assignment $W^{(0)}$ and compute the corresponding Mahalanobis distance $M^{(0)}$. If $M^{(t)}>a$, we randomly switch between one treated unit and one control unit in $W^{(t)}$ to obtain $W^*$ and compute $M^*=M(W^*)$. If $M^*\leq M^{(t)}$, we move to $W^*$, that is, making $t=t+1$ and $W^{(t)}=W^*$. Otherwise, we move to $W^*$ with positive probability $(M^{(t)}/M^*)^{\gamma}$, which prevents us from being trapped in a local optimum. We continue to move until $M^{(t)}\leq a$.
We summarize the whole procedure of PSRR using the Mahalanobis distance in Algorithm \ref{alg:psrr}. 
Tuning parameter $\gamma$ controls the probabilities of movement.  
A larger value of $\gamma$ results in a smaller probability of moving to a less-balanced assignment.
In contrast, a smaller value of $\gamma$ results in more random movements.
In the extreme case of $\gamma=\infty$, we never move to a less-balanced assignment, and we randomly move regardless of the balance if $\gamma=0$.
In our simulation studies and real data analysis, the performance of PSRR is robust for a wide range of values of $\gamma$. The default value is set to $\gamma=10$.

To further improve computational efficiency, we can use the following tricks to compute the Mahalanobis distance:
First, as $X$ is not affected by treatment assignments, we compute $S_{XX}$ only once during the entire procedure of the Fisher randomization tests.
This trick is applicable to rerandomization, GPS, and PSRR.
Second, suppose we switch the $i$th and $j$th elements of $W^{(t)}$ with $W^{(t)}_i=1$ and $W^{(t)}_j=0$ and generate a new assignment $W^*$ with $W^*_i=0$ and $W^*_j=1$.
If we have computed the Mahalanobis distance $M^{(t)}$ of the assignment $W^{(t)}$, we can simplify the calculation of the Mahalanobis distance of $W^*$ by
\begin{equation}
\label{eq:Mstar}
    M(W^*)=M^{(t)} - \Big( 2\sum_{l=1}^n W^{(t)}_l H_{il}-H_{ii} \Big) + \Big( 2\sum_{l=1}^n W^*_l H_{jl}-H_{jj} \Big) + h_i - h_j,
\end{equation}
where $H=XS_{XX}^{-1}X^\T/\{n_t(1-n_t/n)\}$ and $h=(2n_t/n)H\mathbf{1}$ with $\mathbf{1}$ as an $n$-dimensional column vector of 1's. 
The quantities $H$ and $h$ depend only on $X$; thus, we only need to compute them once. The second trick is only applicable to GPS and PSRR. The proof of (\ref{eq:Mstar}) is provided in Web Appendix A.

\begin{algorithm}[t]
\caption{Pair-switching rerandomization}
\label{alg:psrr}
\SetAlgoLined
\textbf{Input} Covariates data $X$, threshold $a$, tuning parameter $\gamma$ (default value $\gamma=10$).

Set $t=0$\;
Set $W^{(0)}$ as $n_t$ elements equal to 1 and $n_c$ elements equal to 0 with random positions\;
Set $M^{(0)}=M(W^{(0)})$\;
\While {$M^{(t)}>a$} {
    Randomly switch the positions of one of the 1's and one of the 0's in $W^{(t)}$ and obtain $W^*$\;
    Set $M^*= M(W^*)$\;
    Sample $J$ from a Bernoulli distribution with probability $\min\{(M^{(t)}/M^*)^{\gamma},1\}$\;
    \If {$J=1$} {
        Set $t=t+1$\;
        Set $W^{(t)}=W^*$\;
        Set $M^{(t)}=M^*$\;
    }
}  
\textbf{Output} $W=W^{(t)}$.
\end{algorithm}

In the design stage, we run Algorithm \ref{alg:psrr} once and output assignment $W^\obs$ to conduct the experiment. In the analysis stage, we run Algorithm \ref{alg:psrr} $B$ times independently and output $\{W^b, b=1,\dots,B\}$. We then follow the procedures in Section \ref{sec:framework} to perform Fisher randomization tests and construct randomization-based confidence intervals. Because we use the same procedure to generate treatment assignments in the design and analysis stages, Fisher randomization tests preserve the significance level, and randomization-based confidence intervals have the desired coverage rate \citep{imbens2015,rosenberger2015rand,luo2021}.

Similar to rerandomization, PSRR still maintains the unbiasedness of the difference-in-means estimator $\widehat\tau$ when $n_t=n_c$. Intuitively, every assignment $W$ has a symmetrical assignment $1 - W$, such that their Mahalanobis distances are equal. Because the distribution of assignments induced by PSRR (Algorithm \ref{alg:psrr}) is driven by the Mahalanobis distance, symmetry implies that $\pr(W_i=1)=\pr(W_i=0)=1/2$. This further implies unbiasedness; see the following Theorem \ref{thm:unbiased}. 

\begin{theorem}
\label{thm:unbiased}
Suppose that $n_t=n_c=n/2$ and $W$ is generated by PSRR. Then, $E( \widehat\tau )=\tau$.
\end{theorem}

If the covariates are related to the outcomes, then the variance of the difference-in-means estimator under PSRR is smaller than that under complete randomization. 
\begin{theorem}
\label{thm:var}
If (i) $n_t=n_c=n/2$, (ii) for $w_i=0,1$, $Y_i(w_i)=\beta_0+\beta^\T X_i+\tau w_i+e_i$, where $\beta_0+\beta^\T X_i$ is the linear projection of $Y_i(0)$ onto $(1, X)$ and $e_i$ is the deviation from the linear projection, and (iii) $\widehat\tau$ and $\overline X_t-\overline X_c$ are normally distributed, then we have
$$
\frac{\Var_{\cre}(\widehat\tau)-\Var_{\psrr}(\widehat\tau)}{\Var_{\cre}(\widehat\tau)}\geq\left(1-a/p\right)R^2,
$$
where the subscripts CR and PSRR represent distributions under complete randomization and pair-switching rerandomization, respectively, and $R^2\equiv\beta^\T\Cov_\cre\left(\overline X_t-\overline X_c\right)\beta/\Var_\cre(\widehat\tau)$.
\end{theorem}

\citet{morgan2012} assumed similar conditions to obtain the variance reduction of rerandomization, which is equal to $\left(1-v_{a,\rr}\right)R^2$ with $v_{a,\rr}=P\left(\chi_{p+2}^{2} \leq a\right)/P\left(\chi_{p}^{2} \leq a\right)$. Because PSRR is more complicated than rerandomization, we only obtain a lower bound for the variance reduction. Compared to $\left(1-v_{a,\rr}\right)R^2$, the bound $\left(1-a/p\right)R^2$ is smaller, but very close to it. For instance, when $p=10$, $R^2=0.5$, and $p_a=0.001$, the former is equal to 44\% and the latter is equal to 42.6\%. Moreover, the lower bound becomes increasingly tight as $p$ increases.

\section{Sequential pair-switching rerandomization}
\label{sec:seq}

In this section, we generalize PSRR to sequentially randomized experiments. We start with a clinical trial that aimed to evaluate the efficacy of the Therapeutic Education System (TES) an internet-delivered treatment for substance or alcohol abuse \citep{campbell2014}. Participant recruitment information was distributed to ten outpatient centers. Patients interested in this project were referred to the researchers for screening. If the patients were eligible, they proceeded to subsequent baseline measurements and randomization. After 15 months, 507 eligible patients were sequentially recruited and randomized immediately upon arrival or within one month. 
Finally, $n_t = 255$ patients were allocated to the treatment group (treatment as usual plus TES), and $n_t = 252$ patients were allocated to the control group (treatment as usual). The primary outcome was the abstinence from drug or heavy alcohol use, assessed using urine drug tests and self-reports. Some baseline covariates were considered to be strong predictors of the outcome, such as the number of days from the participants last drug or alcohol use. In this trial, since the patients received usual therapy regardless of assignment to the treatment or control group, the experimenters could wait for a while (e.g., a month) to recruit a group of patients and perform randomization together.

Similar to the non-sequentially randomized experiment, \citet{zhou2018} proposed a sequential rerandomization (SeqRR) method to balance the baseline covariates. In the following, we first review the basic concepts of sequential rerandomization and then propose a sequential pair-switching rerandomization (SeqPSRR) procedure to reduce the computational cost.

Assume that $n$ patients are divided into $K$ sequential groups of sizes $n_1,\ldots, n_K$. In group $k$ ($k=1,\dots,K$), $n_{tk}=e n_k$ patients are randomly assigned to the treatment group and the remaining $n_{ck}=(1-e)n_k$ patients are assigned to the control group, where $e \in (0,1)$ is the propensity score.
We denote the covariates in the first $k$ groups as $X_{[1:k]}$, whose dimension is $(n_1+\cdots+n_k) \times p$.
We denote the covariance matrix of $X_{[1:k]}$ as $S_{XX[k]}$.
There are two main differences between sequential and classical rerandomizations: (1) when we assign treatment to patients in the former groups, we cannot access covariate data in the latter groups; (2) when we assign treatment to patients in the latter groups, we cannot change the treatment assignment of the patients in the former groups, although we still have to consider covariate balance of all arrived groups.

Sequential rerandomization proceeds as follows:
For the first group, we randomly assign $n_{t1}$ patients to the treatment group and the other $n_{c1}$ to the control group.
This assignment is denoted by vector $W_{[1]}=(W_1,\ldots,W_{n_1})$.
The Mahalanobis distance corresponding to assignment $W_{[1]}$ is defined as follows:
$$
\begin{aligned}
M_1(W_{[1]})&\equiv\left(\overline X_{t[1]}-\overline X_{c[1]}\right)^\T\left[\operatorname{cov}\left(\overline X_{t[1]}-\overline X_{c[1]}\right)\right]^{-1}\left(\overline X_{t[1]}-\overline X_{c[1]}\right) \\ 
&=n_{t1}\left(1-n_{t1}/n_1\right)\left(\overline X_{t[1]}-\overline X_{c[1]}\right)^\T S_{XX[1]}^{-1}\left(\overline X_{t[1]}-\overline X_{c[1]}\right),
\end{aligned}
$$
where $\overline X_{t[1]}=\sum_{i: W_i=1, i\leq n_{1}}X_{i}/n_{t1}$ and $\overline X_{c[1]}=\sum_{i: W_i=0, i\leq n_{1}}X_{i}/n_{c1}$ are the mean vectors of the covariates under treatment and control in the first group, respectively.
If $M_1(W_{[1]})\leq a_1$, a prespecified threshold, we accept the assignment $W_{[1]}$ and conduct the experiment for the patients in the first group. Otherwise, we rerandomize until $M_1(W_{[1]})\leq a_1$ for some $W_{[1]}$. 
If $K=1$, this step is the same as that in the classical rerandomization. 

When the patients in the $k$th group enroll in the experiment,
we randomly assign $n_{tk}$ patients to the treatment group and the other $n_{ck}$ patients to the control group.
This assignment is denoted by a vector $W_{[k]}=(W_{n_{1:(k-1)}+1},\ldots,W_{n_{1:k}})$, where $n_{1:k}=\sum_{l=1}^{k}n_{l}$ is the total number of patients in the first $k$ groups. Then, the Mahalanobis distance corresponding to assignment $W_{[k]}$ is defined as:
$$
\begin{aligned}
M_k(W_{[k]})&\equiv\left(\overline X_{t[1:k]}-\overline X_{c[1:k]}\right)^\T\left[\operatorname{cov}\left(\overline X_{t[1:k]}-\overline X_{c[1:k]}\right)\right]^{-1}\left(\overline X_{t[1:k]}-\overline X_{c[1:k]}\right) \\ 
&=n_{t,1:k}\left(1-n_{t,1:k}/n_{1:k}\right)\left(\overline X_{t[1:k]}-\overline X_{c[1:k]}\right)^\T S_{XX[k]}^{-1}\left(\overline X_{t[1:k]}-\overline X_{c[1:k]}\right),
\end{aligned}
$$
where $n_{t,1:k}=\sum_{l=1}^k n_{tl}$, $n_{c,1:k}=\sum_{l=1}^k n_{cl}$ are the total numbers of patients in the treatment and control arms in the first $k$ groups, and $\overline X_{t[1:k]}=\sum_{i: W_i=1, i\leq n_{1:k}}X_{i}/n_{t,1:k}$, $\overline X_{c[1:k]}=\sum_{i: W_i=0, i\leq n_{1:k}}X_{i}/n_{c,1:k}$ are the mean vectors of the covariates in the treatment and control arms in the first $k$ groups, respectively.
If $M_k(W_{[k]})\leq a_k$, we accept assignment $W_{[k]}$ and conduct the experiment for the patients in the $k$th group. Otherwise, we rerandomize the $n_k$ units in the $k$th group until $M_k(W_{[k]})\leq a_k$ for some $W_{[k]}$.

\citet{zhou2018} provided suggestions on the choice of $a_k$. 
They derived the optimal allocation of the expected number of draws in group $k$, denoted by $s_k$, based on asymptotic arguments (see Proposition 2 therein).
Moreover, they showed that the conditional distribution of $M_k\mid M_{k-1}$ is a non-central chi-square. We can then calculate $a_k$ based on $s_k$ and $M_{k-1}$.

\citet{zhou2018} also showed the unbiasedness of the difference-in-means estimator $\widehat\tau$ for estimating the average treatment effect $\tau$ given $n_{tk}=n_{ck}$, $k=1,\ldots,K$. 
Although a proper asymptotic inference procedure has not been established under sequential rerandomization, Fisher randomization tests are still valid. Like classical rerandomization, the computational cost of Fisher randomization tests under sequential rerandomization is also very high, motivating us to consider sequential pair-switching rerandomization (SeqPSRR).

The SeqPSRR procedure is presented in Algorithm \ref{alg:seqpsrr}. We sequentially replace classical rerandomization with PSRR for each group. When $K=1$, Algorithm \ref{alg:seqpsrr} reduces to Algorithm \ref{alg:psrr}. We also use the tricks discussed in Section \ref{sec:psrr} to compute the Mahalanobis distance. We run Algorithm~\ref{alg:seqpsrr} once and output $W^\obs=(W_{[1]},\ldots,W_{[K]})$ sequentially to conduct the experiment. We sample $B$ assignments $\{W^b,b=1,\dots,B\}$ following Algorithm~\ref{alg:seqpsrr} independently, perform Fisher randomization tests, and construct randomization-based confidence intervals following the procedures introduced in Section \ref{sec:framework}.

\begin{algorithm}[!ht]
\caption{Sequential pair-switching rerandomization}
\label{alg:seqpsrr}
\SetAlgoLined
\textbf{Input} Expected numbers of draws in each group, $s_1, \ldots, s_K$, and tuning parameter $\gamma$ (default value $\gamma=10$).

Set $M_{[0]}=0$\;
Set $n_0=0$\;

\For {$k=1,\ldots,K$} {
    \textbf{Input} Covariates data $X_{[1:k]}$.

    Set $a_k=n_k(n_{1:k})^{-1}q_k $, where $q_k$ is the lower $1/s_k$ quantile of a non-central chi-square distribution with $p$ degrees of freedom and a non-central parameter $ n_{1:(k-1)} (n_k)^{-1} M_{[k-1]}$\;
    Set $t=0$\;
    Set $W^{(0)}_{[k]}$ as $n_{tk}$ elements equal to 1 and $n_{ck}$ elements equal to 0 with random positions\;
    Set $M^{(0)}_{[k]}=M_k(W^{(0)}_{[k]})$\;
    \While {$M^{(t)}_{[k]}>a_k$} {
        Randomly switch the positions of one of the 1's and one of the 0's in $W^{(t)}_{[k]}$ and obtain $W^*_{[k]}$\;
        Set $M^*_{[k]}= M_k(W^*_{[k]})$\;
        Sample $J$ from a Bernoulli distribution with probability $\min\left\{\left(M^{(t)}_{[k]}/M^*_{[k]}\right)^{\gamma},1\right\}$\;
        \If {$J=1$} {
            Set $t=t+1$\;
            Set $W^{(t)}_{[k]}={W}^*_{[k]}$\;
            Set $M^{(t)}_{[k]}=M^*_{[k]}$\;
        }
    }  
    Set $M_{[k]}=M_{[k]}^{(t)}$\;
    \textbf{Output} $W_{[k]}=W^{(t)}_{[k]}$.
}
\textbf{Output} $W=(W_{[1]},\ldots,W_{[K]})$.
\end{algorithm}

By applying the arguments in Theorem \ref{thm:unbiased} to each group separately, we show that the SeqPSRR also maintains the unbiasedness of the difference-in-means estimator when $n_{tk} = n_{ck}$ for all $k=1,\dots,K$. 
\begin{theorem}
\label{thm:seq_unbiased}
Suppose that $n_{tk}=n_{ck}=n_k/2, k=1,\ldots,K$, and $W$ is generated from SeqPSRR. Then, $E(\widehat\tau)=\tau$.
\end{theorem}

\citet{zhou2018} showed that, compared to complete randomization, the proportion of variance reduction of sequential rerandomization is $\left\{1-E(M_K)/p\right\}R^2$. Under the more complicated SeqPSRR, we can obtain a lower bound for the proportion of variance reduction. Following \citet{morgan2012} and \citet{zhou2018}, we assume equal treatment group sizes, additive treatment effects, and normally distributed difference-in-means of the covariates and outcomes.
\begin{theorem}
\label{thm:seq_var}
If (i) $n_{tk}=n_{ck}=n_k/2, k=1,\ldots,K$, (ii) for $w_i=0,1$, $Y_i(w_i)=\beta_0+\beta^\T X_i+\tau w_i+e_i$, where $\beta_0+\beta^\T X_i$ is the linear projection of $Y_i(0)$ onto $(1, X)$ and $e_i$ is the deviation from the linear projection, and (iii) $\widehat\tau$ and $\overline X_t-\overline X_c$ are normally distributed, then we have 
$$
\frac{\Var_{\cre}(\widehat\tau)-\Var_{\seqpsrr}(\widehat\tau)}{\Var_{\cre}(\widehat\tau)}\geq\left(1-a_K/p\right)R^2,
$$ 
where the subscript SeqPSRR represents the distribution under sequential pair-switching rerandomization.
\end{theorem}

\section{Simulation studies}
\label{sec:sim}

\subsection{Non-sequentially randomized experiments}
\label{sec:sim_nonseq}

We compare PSRR with existing methods in both non-sequentially and sequentially randomized experiments. To fairly compare the speed of these methods, we use R to implement all methods. For each case, we run the code on an Intel Xeon E5-2690 V4 processor (2.6GHz, 35M Cache, 28 Core, 128G Memory). We replicate the simulation $n_{\rm rep}=1000$ times to examine the repeated sampling properties.

In this section, we consider non-sequentially randomized experiments. The covariates are generated from the standard normal distribution, $X_{i j}\stackrel{i.i.d.}\sim N(0,1)$, $i=1,\dots,n, \ j=1,\dots,p$, where i.i.d. stands for ``independent and identically distributed.'' Similar to the simulation setups in \citet{johansson2021}, the potential outcomes $Y_i(0)$, $i=1,\dots,n$, are generated independently by a linear regression model, $Y_{i}(0)=X_{i 1}+\cdots+X_{i p}+\epsilon_{i}$, where $\epsilon_{i}\stackrel{i.i.d.}\sim N(0,\sigma^2_{\epsilon})$. We set $n=30,50,100$, $p=10$, and choose $\sigma^2_{\epsilon}$ such that $R^2=\Var(X_{i1}+\cdots+X_{ip})/\{\Var(X_{i1}+\cdots+X_{ip})+\sigma^2_{\epsilon}\}=0.2$ or 0.5. The $R^2$ measures the correlations between covariates and potential outcomes. Usually, the larger the R-square, the greater the benefit of balancing covariates \citep{morgan2012}. To examine the size (type one error) and power (one minus type two error) of the Fisher randomization tests, we respectively set $Y_i(1)=Y_i(0)$ and $Y_{i}(1)=Y_{i}(0)+0.3 \sqrt{\Var\{ Y_i(0) \}}$. Both covariates and potential outcomes are generated once and then kept fixed. The results for larger sample sizes, $n=500,1000,2000$, are similar and provided in Web Appendix C.

We consider four design methods: complete randomization (CR), greedy pair-switching (GPS, \citet{krieger2019bka}), rerandomization (RR, \citet{morgan2012}), and pair-switching rerandomization (PSRR). We consider equal-sized treatment and control groups. For the two rerandomization methods, the threshold $a$ of the Mahalanobis distance satisfies $p_a=0.001$, following the recommendations of \citet{li2018}. For PSRR, we set the default value of the tuning parameter $\gamma$ to 10. We also examine the performance of PSRR with different values of $\gamma$. The results are similar and provided in Web Appendix C.
We use three randomness metrics to measure the randomness of these methods, see Web Appendix B.
We perform Fisher randomization tests with a significance level $\alpha=0.05$ and construct randomization-based confidence intervals with a nominal coverage rate of 95\%. We set $B=1000$. We use both the proposed exact approach and bisection method \citep{wang2020ci,luo2021} to invert the Fisher randomization tests to confidence intervals and compare their performances.

\begin{table}[p]

\caption{\label{tab:sim_main}Statistical and computational performance of different methods when $R^2=0.5$.}
\centering
\begin{threeparttable}
\begin{tabular}[t]{lrrrrrrrrr}
\toprule
\multicolumn{1}{c}{ } & \multicolumn{6}{c}{Inference ($\times 10^{-2}$)} & \multicolumn{3}{c}{Run time (second)} \\
\cmidrule(l{3pt}r{3pt}){2-7} \cmidrule(l{3pt}r{3pt}){8-10}
\multicolumn{1}{c}{Method} & \multicolumn{1}{c}{Bias} & \multicolumn{1}{c}{SD} & \multicolumn{1}{c}{Size} & \multicolumn{1}{c}{Power} & \multicolumn{1}{c}{CP} & \multicolumn{1}{c}{Length} & \multicolumn{1}{c}{Sample} & \multicolumn{1}{c}{Exact} & \multicolumn{1}{c}{Bisection}\\
\midrule
\addlinespace[0.3em]
\multicolumn{10}{c}{\textbf{Non-sequential: $n=30$}}\\
\hspace{1em}CR & 2.6 & 118 & 5.3 & 11 & 94.7 & 490 & 0.2 & 0.02 & 0.61\\
\hspace{1em}GPS & 1.5 & 88 & 4.3 & 17 & 95.7 & 381 & 18.5 & 0.02 & 0.61\\
\hspace{1em}RR & 2.2 & 92 & 4.6 & 17 & 95.4 & 389 & 120.1 & 0.02 & 0.62\\
\hspace{1em}PSRR & 2.2 & 92 & 4.2 & 15 & 95.8 & 392 & 5.2 & 0.02 & 0.61\\
\midrule
\addlinespace[0.3em]
\multicolumn{10}{c}{\textbf{Non-sequential: $n=50$}}\\
\hspace{1em}CR & 3.2 & 103 & 4.2 & 16 & 95.8 & 420 & 0.2 & 0.02 & 0.58\\
\hspace{1em}GPS & 3.2 & 76 & 4.9 & 30 & 95.1 & 308 & 64.9 & 0.02 & 0.59\\
\hspace{1em}RR & 1.6 & 77 & 3.8 & 26 & 96.2 & 320 & 83.6 & 0.02 & 0.58\\
\hspace{1em}PSRR & 2.1 & 75 & 4.0 & 24 & 96.0 & 319 & 3.6 & 0.02 & 0.57\\
\midrule
\addlinespace[0.3em]
\multicolumn{10}{c}{\textbf{Non-sequential: $n=100$}}\\
\hspace{1em}CR & 1.5 & 89 & 6.2 & 31 & 93.8 & 349 & 0.2 & 0.03 & 0.60\\
\hspace{1em}GPS & 1.0 & 62 & 4.8 & 55 & 95.2 & 252 & 351.0 & 0.03 & 0.60\\
\hspace{1em}RR & 0.1 & 68 & 6.2 & 50 & 93.8 & 265 & 69.8 & 0.03 & 0.60\\
\hspace{1em}PSRR & 0.2 & 69 & 4.9 & 50 & 95.1 & 265 & 3.2 & 0.03 & 0.60\\
\midrule
\addlinespace[0.3em]
\multicolumn{10}{c}{\textbf{Sequential: $K=3, n_k=20$}}\\
\hspace{1em}SeqCR & 2.6 & 99 & 5.3 & 18 & 94.7 & 406 & 0.4 & 0.02 & 0.58\\
\hspace{1em}SeqRR & 1.7 & 74 & 5.6 & 32 & 94.4 & 302 & 108.8 & 0.02 & 0.58\\
\hspace{1em}SeqPSRR & 1.2 & 75 & 5.0 & 33 & 95.0 & 302 & 16.6 & 0.02 & 0.57\\
\midrule
\addlinespace[0.3em]
\multicolumn{10}{c}{\textbf{Sequential: $K=5, n_k=20$}}\\
\hspace{1em}SeqCR & 1.1 & 85 & 4.6 & 32 & 95.4 & 340 & 0.6 & 0.02 & 0.63\\
\hspace{1em}SeqRR & 0.4 & 58 & 4.2 & 57 & 95.8 & 240 & 94.8 & 0.02 & 0.63\\
\hspace{1em}SeqPSRR & 1.9 & 59 & 4.3 & 57 & 95.7 & 239 & 18.4 & 0.02 & 0.63\\
\midrule
\addlinespace[0.3em]
\multicolumn{10}{c}{\textbf{Sequential: $K=10, n_k=20$}}\\
\hspace{1em}SeqCR & 0.7 & 63 & 6.1 & 55 & 93.9 & 250 & 1.3 & 0.03 & 0.78\\
\hspace{1em}SeqRR & 0.7 & 46 & 5.7 & 83 & 94.3 & 180 & 86.5 & 0.03 & 0.78\\
\hspace{1em}SeqPSRR & 0.8 & 46 & 4.3 & 84 & 95.7 & 179 & 26.6 & 0.03 & 0.77\\
\bottomrule
\end{tabular}
\begin{tablenotes}
\item Note: Bias, absolute bias; SD, standard deviation; CP, coverage probability; Length, mean interval length; Sample, sampling $10^3$ acceptable assignments; Exact/Bisection, determining interval endpoints by the exact approach/bisection method.
\end{tablenotes}
\end{threeparttable}
\end{table}

\begin{figure}[t]
\centering
    \includegraphics[scale = 0.85]{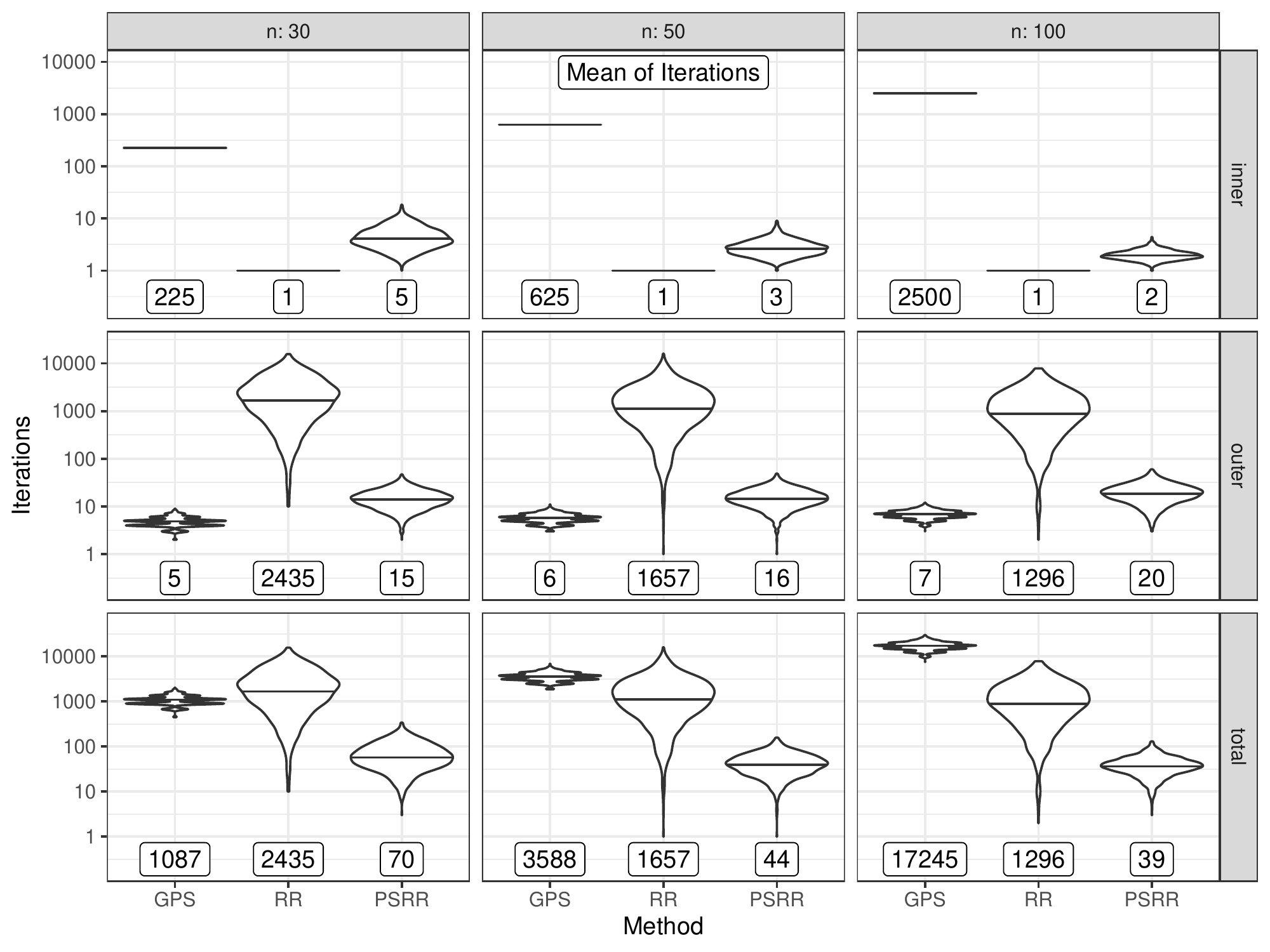}
\caption{Numbers of inner, outer, and total iterations of GPS, RR, and PSRR needed to find one acceptable assignment in the non-sequential setting when $n=30,50,100$, respectively.}
\label{fig:sim_ns_iter}
\end{figure}

Table \ref{tab:sim_main} shows the results of the non-sequentially randomized experiments when $R^2=0.5$. The results for $R^2=0.2$ are provided in Web Appendix C. 
First, the difference-in-means estimator $\widehat \tau$ under all design methods has negligible finite-sample biases.
Second, compared with CR, the other three design methods reduce the standard deviation of $\widehat \tau$ by 25\%--31\%, 22\%--25\%, and 22\%--27\%, respectively. Notably, compared with RR and PSRR, although GPS improves the balance of assignments, it does not significantly reduce the variance of $\widehat \tau$. This is because when the Mahalanobis distance is already small, the gain in efficiency by further pursuing balance is often negligible \citep{johansson2021}.
Third, under all design methods and sample sizes, Fisher randomization tests control the type one error, and randomization-based confidence intervals reach the nominal coverage rate. 
Fourth, compared to CR, the other three design methods increase the power by 57\%--91\%, 60\%--68\%, and 38\%--63\%, respectively, and reduce the interval length by 22\%--28\%, 21\%--24\%, and 20\%--24\%, respectively. Because the bisection method produces almost the same (but slightly wider) interval as the exact approach, we do not present its coverage probability and interval length owing to space restrictions.
Fifth, PSRR dramatically reduces the computational costs: it is 4--110 times faster than GPS and 22--23 times faster than RR. Moreover, with an increase in $n$, the computation time of GPS increases dramatically. In contrast, the computation times of RR and PSRR decrease as $n$ increases because the number of iterations used to find each acceptable assignment decreases, as shown in Figure \ref{fig:sim_ns_iter}. The improvement in the speed of PSRR relative to RR is not strongly dependent on the sample size. Finally, compared to the computational time required to generate the desired assignments, the time required to construct confidence intervals using the proposed exact approach is negligible (approximately 0.02 seconds), while the computational time of the bisection method is non-negligible (approximately 0.6 seconds) under both CR and PSRR.

Figure \ref{fig:sim_ns_iter} shows that PSRR requires much smaller \textit{total} iterations than GPS and RR to find an acceptable assignment, which is why PSRR is much faster than the other two methods. Because the total number of iterations is equal to the average number of \textit{inner} iterations (the number of attempts before jumping to a new candidate assignment) times the number of \textit{outer} iterations (the number of candidate assignments before finding an acceptable assignment), we further compare the inner and outer iterations of these three methods.
Because GPS enumerates all pair-switched assignments of the current assignment and jumps to the best, its inner iterations are equal to $(n/2)^2$, which increases rapidly as $n$ increases. It takes approximately 5--7 outer iterations to find an acceptable assignment (i.e., to reach a local optimum).
RR has no inner loop and requires approximately 1296--2435 outer iterations to find one acceptable assignment. 
PSRR takes approximately 2--5 attempts to jump to a new candidate assignment, which is much smaller than that of GPS, and 15--20 outer iterations to find one acceptable assignment, which is much smaller than that of RR.
Overall, PSRR outperforms both GPS and RR in terms of the number of total iterations (39--70 compared to 1087--17245 and 1296--2435, respectively), which leads to PSRR sampling acceptable assignments more quickly.

\subsection{Sequentially randomized experiments}
\label{sec:sim_seq}

In this section, we consider sequentially randomized experiments. The experimental units are recruited in $K=3,5,10$ groups, and each group has $n_k=20$ units. We also consider the setting of $n_k=50$ and provide the results in Web Appendix C. We simulate the covariates and potential outcomes in the same way as those in Section \ref{sec:sim_nonseq}.
We consider three design methods: sequential complete randomization (SeqCR), sequential rerandomization (SeqRR), and sequential pair-switching rerandomization (SeqPSRR). For the two rerandomization methods, we set $S=1000$ and determine $s_k$ and $a_k$ following the strategy proposed by \citet{zhou2018}. Specifically, we set $(s_1,s_2,s_3)=(30, 136, 834)$, when $K=3$; $(s_1,\ldots,s_5)=(10, 10, 29, 133, 818)$ when $K=5$; and $(s_1,\ldots,s_{10})=(10, 10, 10, 10, 10, 10, 10, 28, 128, 774)$, when $K=10$. Because $a_k$ is determined based on chi-square approximation, to find an assignment such that $M_k\leq a_k$, the actual number of rerandomizations is probably much greater than $s_k$ in the finite sample. We follow \citet{zhou2018} to allow at most $10s_k$ rerandomizations and output the best assignment if all these rerandomizations are not acceptable. We further examine the influence of the $10s_k$ constraint in Web Appendix C. 

Table \ref{tab:sim_main} presents the performance of all methods when $R^2=0.5$. The results for $R^2=0.2$ are provided in Web Appendix C. The overall conclusions are similar to those in the non-sequentially randomized experiments. Under all three designs, the biases of $\widehat{\tau}$ are negligible, Fisher randomization tests control for type one error, and randomization-based confidence intervals have the desired coverage rate. Compared with SeqCR, the other two methods, SeqRR and SeqPSRR, reduce the standard deviation of the difference-in-means estimator by 25\%--33\%, increase the power by 50\%--82\%, and reduce the interval length by 26\%--30\%. SeqPSRR is 3--7 times faster than SeqRR. 
The improvement in speed decreases as $K$ increases, because as $K$ becomes larger, the $s_k$'s become smaller in most groups, and smaller $s_k$ values directly reduce the improvement in speed (in the extreme case with $s_k=1$, SeqCR, SeqRR, and SeqPSRR are equivalent in group $k$).
Overall, SeqPSRR achieves comparable precision of point and interval estimation and power of tests, with less computational cost than SeqRR.

\section{Clinical trial examples}
\label{sec:real}

\subsection{A phase 1 clinical trial}

We revisit the non-sequentially randomized experiment introduced in Section \ref{sec:framework} and illustrate the applicability and advantages of PSRR. In this experiment, $n_t=20$ and $n_c=10$. To balance the eight important covariates using RR and PSRR, we set the threshold $a$ as the $0.001$ quantile of $\chi^2_8$, which leads to $a=0.86$. In contrast, the Mahalanobis distance corresponding to the actual assignment is 10.95, the 0.795 quantile of  $\chi^2_8$, which is a rather unbalanced assignment.

\begin{figure}[p]
\centering
    \includegraphics[scale = 0.68]{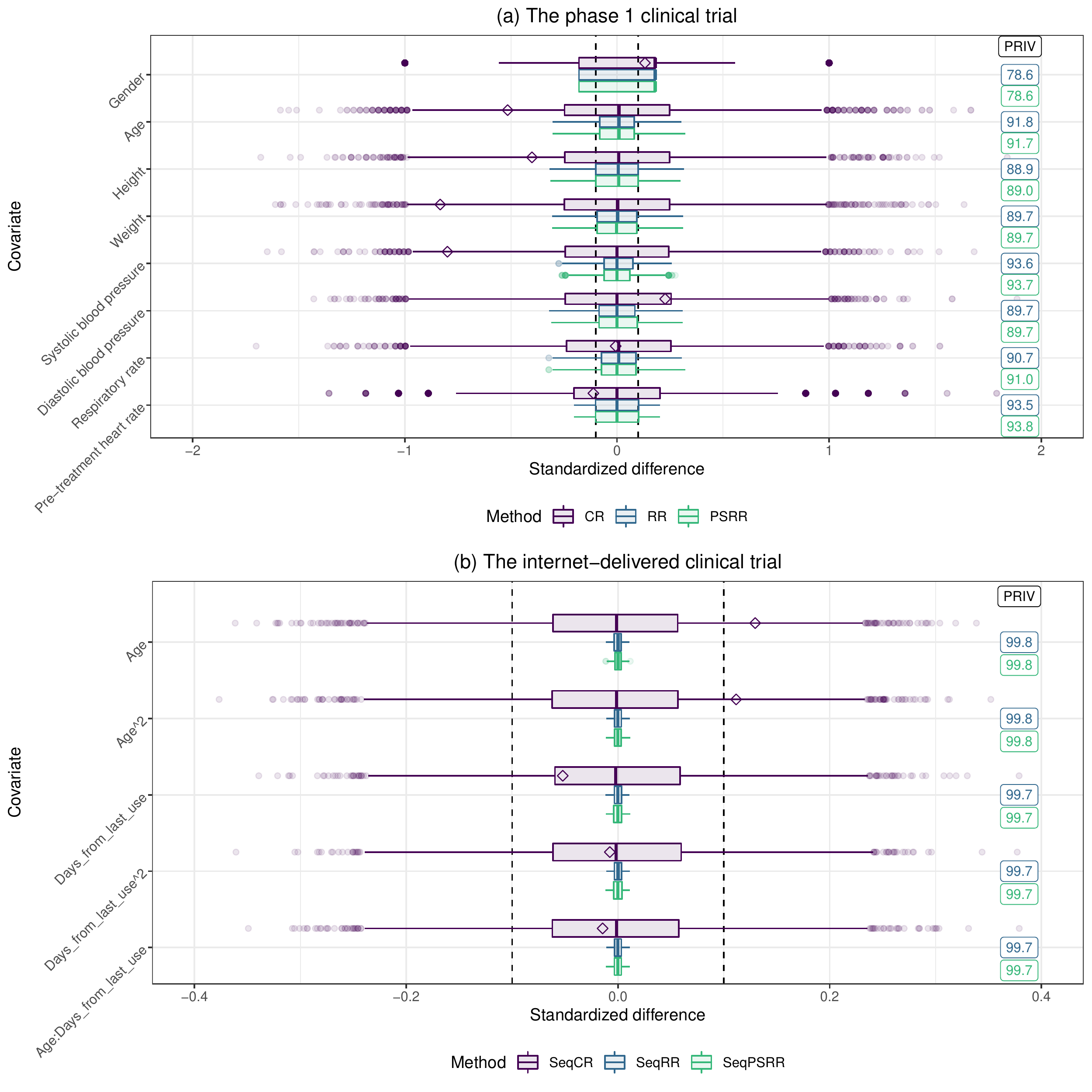}
\caption{Box-plot of standardized differences in covariate means for two datasets. The dashed lines indicate the recommended univariate balance thresholds $[-0.1,0.1]$ \citep{austin2009balance}. The diamonds indicate the standardized differences in covariate means for the actual assignment in the experiment. PRIV stands for the empirical percent reductions in variance. For each covariate, the methods from top to bottom are CR, RR, and PSRR for sub-plot (a) and SeqCR, SeqRR, and SeqPSRR for sub-plot (b), respectively. This figure appears in color in the electronic version of this article, and any mention of color refers to that version.}
\label{fig:rd_dif}
\end{figure}

We generate 10,000 assignments by CR, RR, and PSRR. RR takes approximately 19 minutes, whereas PSRR takes approximately one minute. Figure \ref{fig:rd_dif}(a) shows the empirical distributions of the standardized differences in covariate means and the empirical percent reductions in variance (PRIVs) relative to CR. PSRR performs similarly to RR in terms of balancing covariates. The empirical PRIVs for the covariates are close to the theoretical lower bound $100(1-a/p)\%\approx 89.3\%$. According to Theorem \ref{thm:var}, the lower bound of the PRIV for the treatment effect estimation is $\{100(1-a/p)R^2\}\%\approx 38.6\%$, where $R^2=0.432$ is the adjusted R-square obtained by regressing the outcome on the covariates. To examine this conclusion, a simulation based on semi-synthetic data is provided in Web Appendix C. If we use the assignment generated by PSRR to replace the original assignment in the experiment, all covariates will be well-balanced in the design stage. Thus, the difference in post-treatment heart rate will no longer be attributable to the baseline difference.

\subsection{An internet-delivered clinical trial}

We consider the sequentially randomized experiment introduced in Section \ref{sec:seq} and show the applicability and advantages of SeqPSRR. We split $n=507$ participants into $K=15$ sequential groups (with an approximate waiting time of one month per group) with sizes $n_1=31$ and $n_k=34$, $ 2 \leq k \leq 15$. We set $n_{tk}=17$, $1\leq k\leq 15$, such that $n_t=255$. To balance the covariates of age, the number of days since the last use of drugs or alcohol, and their quadratic terms and interaction ($p=5$) using SeqRR and SeqPSRR, we set $s_k=10$ for $1 \leq k\leq 12$, $s_{13}=12$, $s_{14}=68$, and $s_{15}=800$.

We generate 10,000 assignments by SeqCR, SeqRR, and SeqPSRR. SeqPSRR is more than twice as fast as SeqRR (334 seconds versus 867 seconds). Figure \ref{fig:rd_dif}(b) shows the empirical distributions of the standardized differences in the covariate means and PRIVs relative to SeqCR. For the original assignment, the standardized differences in age and their quadratic terms are larger than 0.1, indicating that they are not well-balanced \citep{austin2009balance}. In contrast, both SeqRR and SeqPSRR produce balanced assignments. The empirical PRIVs for the covariates are close to the theoretical lower bound $100(1-a_K/p)\%\approx 99.6\%$. According to Theorem \ref{thm:seq_var}, the lower bound of the PRIV for the treatment effect estimation is $\{100(1-a_K/p)R^2\}\%\approx 13.8\%$, where $R^2=0.139$ is the adjusted R-square obtained by regressing the outcome on the covariates. To examine this conclusion, a simulation based on semi-synthetic data is provided in Web Appendix C. Overall, SeqPSRR balances the covariates as well as SeqRR and reduces the computational cost of conducting Fisher randomization tests.

\section{Discussion}
\label{sec:dis}

Rerandomization can improve the efficiency of statistical inference by balancing baseline covariates in the design stage. However, the low sampling efficiency of classical rerandomization forces researchers to make a trade-off between feasibility and covariate balance, which leads to inferior statistical performance. In this article, we propose PSRR and SeqPSRR to balance the baseline covariates. Compared with classical rerandomization and sequential rerandomization, the proposed methods can achieve comparable precision of point and interval estimates and power of tests, but with a much lower computational cost. We derive the unbiasedness and a lower bound for the variance reduction of the difference-in-means estimator under both PSRR and SeqPSRR. In addition, we propose an exact approach to invert Fisher randomization tests to construct randomization-based confidence intervals. Extensive simulation studies and two clinical trial data analyses demonstrate the advantages of the proposed methods. 
Under PSRR, the assignments are no longer uniformly distributed on $\mathcal{W}_a$. It is challenging to derive the (asymptotic) distribution of commonly used test statistics, such as the difference-in-means estimator. We leave this problem to future work. Another limitation of the proposed methods is that they are not applicable to fully sequential experiments, in which units enroll the experiments one by one.

In the clinical trial introduced in Section \ref{sec:framework}, baseline measurements were performed on days 0 and 1, randomization was performed on day 3, and actual treatment allocation was conducted on day 4.
During this period, some baseline covariates, such as blood pressure, may change. However, Fisher randomization tests remain valid as long as the same covariates are used in the design and analysis stages. In short, we emphasize that ``one should analyze as one designs'' \citep[][Section 6.4]{rosenberger2015rand}. 
Since earlier values of covariates are often as predictive of outcomes as later values of covariates, balancing earlier measured covariates can also improve statistical efficiency. 
If we take the values that are the closest to the treatment allocation as the true values of the covariates, we can view earlier values of the covariates as measurements with errors. \citet{wang2021impact} studied the impact of measurement error on covariate-adaptive randomization. It would be interesting to extend their theory to PSRR.

In the clinical trial introduced in Section \ref{sec:seq}, 49 out of 255 participants in the TES group were reported to have not completed the entire 12-week TES course, whereas 252 participants in the control group did not have access to TES during the study.
For this one-sided noncompliance (treatment switching) issue, our method provides a valid intention-to-treat (ITT) analysis of the effect of \textit{assignment} \citep[][Chapter 23]{imbens2015}. To infer the effect of the actual \textit{receipt of treatment}, \citet{mattei2020assessing} addressed the problem of treatment switching using principal stratification \citep{frangakis2002principal} and proposed a Bayesian approach. \citet{rubin1998more} established a framework for utilizing Fisher randomization tests in the presence of imperfect compliance. We can adopt Rubin's framework to handle noncompliance problems in PSRR.

Rerandomization balances covariates in the design stage. Another approach to dealing with covariate imbalance is to use regression adjustment in the analysis stage \citep{lin2013agnostic,bloniarz2016lasso,liu2020regression,lei2021regression,su2021model}, which has also been combined with rerandomization to further improve efficiency \citep{li2020re}. It would be interesting to combine the PSRR and regression-adjusted Fisher randomization tests \citep{zhao2021}.

Rerandomization has been extended to experiments with multiple arms \citep{branson2016, li2020fa}, where we need to balance multiple contrasts of covariate means simultaneously. Thus, conducting Fisher randomization tests might face a more severe computational burden than rerandomization in two-arm experiments. Therefore, it would be interesting to generalize PSRR to experiments with multiple arms, including the factorial experiments.

We use the Mahalanobis distance as a balance measure. It is straightforward to extend our methods to rerandomization using other balance measures, such as the Mahalanobis distance within tiers of covariate importance \citep{morgan2015}, rank-based balance measure with estimated weights of the covariates \citep{johansson2020}, ridge rerandomization \citep{branson2021}, and PCA rerandomization \citep{zhang2021}. 

Recently, the choice of threshold $a$ in rerandomization has been further investigated. \citet{kapelner2019} proposed a procedure to determine the optimal rerandomization threshold based on a trade-off between the observed imbalance and the risk of unobserved imbalance. \citet{banerjee2020} provided guidelines for choosing rerandomization thresholds based on a trade-off between covariate balance and robustness.
In practice, classical rerandomization using these recommended thresholds may not be computationally feasible. PSRR addresses this issue.

\section*{Acknowledgements}
The authors are grateful to the Associate Editor and two referees for their valuable comments. 
Dr. Ke Zhu would like to thank Professors Donald Rubin and Ke Deng for their helpful comments during the causal inference seminar at Tsinghua University. Dr. Hanzhong Liu's research is supported by the National Natural Science Foundation of China (Grant No. 12071242) and Guo Qiang Institute of Tsinghua University.
The information reported here results from secondary analyses of data from clinical trials conducted by the National Institute on Drug Abuse (NIDA). Specifically, data from NIDA-CPU-0006 (A Phase 1 Parallel-Group, Double-Blind, Placebo-Controlled Cardiovascular and Behavioral Study Assessing Interactions Between Single Doses of Oral Reserpine and Intravenous Methamphetamine) and NIDA-CTN-0044 (Web-delivery of Evidence-Based, Psychosocial Treatment for Substance Use Disorders) were included. NIDA databases and information are available at http://datashare.nida.nih.gov.

\section*{Data Availability Statement}

The data that support the findings in this paper are openly available in National Institute on Drug Abuse (NIDA) databases at https://datashare.nida.nih.gov. Specifically, data from \citet{jones2017} and \citet{nunes2014} were included.

\bibliographystyle{apalike}
\bibliography{causal.bib}

\newpage

\section*{Supporting Information}
	
\appendix

Appendix \ref{sec:proof} provides proofs of theoretical results in the main text. 
Appendix \ref{sec:rand} introduces three randomness metrics proposed in \citet{krieger2019bka}.
Appendix \ref{sec:add_num} reports additional simulation results.

\section{Proof of main results}
\label{sec:proof}

\setcounter{equation}{0}
\renewcommand{\theequation}{A\arabic{equation}}

\subsection{Proof of Theorem~\ref{thm:ci}}

\begin{proof}

For $\Wrep\in\mathcal{W}_a$ and $\Wrep\neq W^\obs$, we first solve the equation $\widehat{\tau}(\mathbf{Y}_{\htau}^{\imp},\Wrep)=\widehat{\tau}(\mathbf{Y}_{\htau}^{\imp},W^\obs)$. Under the sharp null $H_0^\htau$, the imputed outcomes are
$$
\begin{aligned}
&\widetilde{Y}_{i}(1)= \begin{cases}{Y}_{i}(1) & \text { if } W_{i}=1, \\
Y_i(0)+\htau &\text { if } W_{i}=0, \end{cases} \\
&\widetilde{Y}_{i}(0)= \begin{cases}{Y}_{i}(0) & \text { if } W_{i}=0, \\
Y_i(1)-\htau &\text { if } W_{i}=1. \end{cases} 
\end{aligned}
$$
By comparing $\Wrep$ and $W^\obs$, we denote $n_{\mathrm{dif}}\equiv\sum_{i=1}^n I_{\{W_i^\obs=1,\Wrep_i=0\}}=\sum_{i=1}^n I_{\{W_i^\obs=0,\Wrep_i=1\}}$. For the observed assignment $W^\obs$, we have 
$$
\widehat{\tau}^\obs\equiv\widehat{\tau}(\mathbf{Y}_\htau^{\imp},W^\obs)=\frac{1}{n_1}\sum_{W_i^\obs=1}{Y}_{i}(1)-\frac{1}{n_0}\sum_{W_i^\obs=0}{Y}_{i}(0).
$$
For each hypothetical assignment $\Wrep$, we have
$$
\begin{aligned}
\widehat{\tau}(\mathbf{Y}_\htau^{\imp},\Wrep)=&\frac{1}{n_1}\sum_{\Wrep_i=1}\widetilde{Y}_{i}(1)-\frac{1}{n_0}\sum_{\Wrep_i=0}\widetilde{Y}_{i}(0) \\
=&\frac{1}{n_1}\left\{\sum_{W_i^\obs=1,\Wrep_i=1}{Y}_{i}(1)+\sum_{W_i^\obs=0,\Wrep_i=1}{Y}_{i}(0)+n_{\mathrm{dif}}\htau\right\}\\
&-\frac{1}{n_0}\left\{\sum_{W_i^\obs=0,\Wrep_i=0}{Y}_{i}(0)+\sum_{W_i^\obs=1,\Wrep_i=0}{Y}_{i}(1)-n_{\mathrm{dif}}\htau\right\}.
\end{aligned}
$$
Then, we have
$$
\begin{aligned}
&\widehat{\tau}(\mathbf{Y}_{\htau}^{\imp},\Wrep)=\widehat{\tau}(\mathbf{Y}_{\htau}^{\imp},W^\obs) \\
\Leftrightarrow&
\frac{1}{n_1}\left\{\sum_{W_i^\obs=1,\Wrep_i=1}{Y}_{i}(1)+\sum_{W_i^\obs=0,\Wrep_i=1}{Y}_{i}(0)+n_{\mathrm{dif}}\htau\right\} \\
&-\frac{1}{n_0}\left\{\sum_{W_i^\obs=0,\Wrep_i=0}{Y}_{i}(0)+\sum_{W_i^\obs=1,\Wrep_i=0}{Y}_{i}(1)-n_{\mathrm{dif}}\htau\right\}
=\frac{1}{n_1}\sum_{W_i^\obs=1}{Y}_{i}(1)-\frac{1}{n_0}\sum_{W_i^\obs=0}{Y}_{i}(0) \\
\Leftrightarrow&
\left(\frac{1}{n_1}+\frac{1}{n_0}\right)n_{\mathrm{dif}}\htau =\frac{1}{n_1}\left\{\sum_{W_i^\obs=1,\Wrep_i=0}{Y}_{i}(1)-\sum_{W_i^\obs=0,\Wrep_i=1}{Y}_{i}(0)\right\}\\
&-\frac{1}{n_0}\left\{\sum_{W_i^\obs=0,\Wrep_i=1}{Y}_{i}(0)-\sum_{W_i^\obs=1,\Wrep_i=0}{Y}_{i}(1)\right\}\\
\Leftrightarrow&
\left(\frac{1}{n_1}+\frac{1}{n_0}\right)n_{\mathrm{dif}}\htau
=\left(\frac{1}{n_1}+\frac{1}{n_0}\right)\left\{\sum_{W_i^\obs=1,\Wrep_i=0}{Y}_{i}(1)-\sum_{W_i^\obs=0,\Wrep_i=1}{Y}_{i}(0)\right\}.
\end{aligned}
$$
Thus, we have 
$$
\htau_b=\frac{\sum_{W_i^\obs=1,\Wrep_i=0}Y_i(1)- \sum_{W_i^\obs=0,\Wrep_i=1}Y_i(0)}{n_{\mathrm{dif}}}.
$$

Next, we prove that $\htau_{(\lfloor \alpha \cardW \rfloor +1)}=\sup\left\{\htau: p(\htau)\leq \alpha\right\}\equiv\tl$. According to the definition of supremum, we only need to show (i) for any $\htau\in\left\{\htau: p(\htau)\leq \alpha\right\}$, $\htau\leq\htau_{(\lfloor \alpha \cardW \rfloor +1)}$ and (ii) for any $\htau<\htau_{(\lfloor \alpha \cardW \rfloor +1)}$, $\htau\in\left\{\htau: p(\htau)\leq \alpha\right\}$.

(i) When $\htau=\htau_{(\lfloor \alpha \cardW \rfloor +1)}\geq\htau_{(\lfloor \alpha \cardW \rfloor +1)}\geq...\geq\htau_{(1)}$, there are \textit{at least} $\lfloor \alpha \cardW \rfloor +1$ hypothetical assignments $\Wrep$ such that $\widehat{\tau}(\mathbf{Y}_\htau^{\imp},\Wrep)\geq\widehat{\tau}^\obs$ (not exactly $\lfloor \alpha \cardW \rfloor +1$ since $\htau_{(\lfloor \alpha \cardW \rfloor +2)},\htau_{(\lfloor \alpha \cardW \rfloor +3)},...$ may equal to $\htau_{(\lfloor \alpha \cardW \rfloor +1)}$). Thus, we have
$$
p(\htau_{(\lfloor \alpha \cardW \rfloor +1)})\geq \frac{\lfloor \alpha \cardW \rfloor +1}{\cardW}=\frac{ \alpha \cardW-( \alpha \cardW-\lfloor \alpha \cardW \rfloor) +1}{\cardW}>\frac{ \alpha \cardW}{\cardW}=\alpha,
$$
where the last inequality is due to $\alpha \cardW-\lfloor \alpha \cardW \rfloor<1$.
Since $p(\htau)$ is non-decreasing, for any $\htau\in\left\{\htau: p(\htau)\leq \alpha\right\}$,
$$
p(\htau)\leq \alpha\leq p(\htau_{(\lfloor \alpha \cardW \rfloor +1)})\Rightarrow \htau\leq\htau_{(\lfloor \alpha \cardW \rfloor +1)}.
$$

(ii) For any $\htau<\htau_{(\lfloor \alpha \cardW \rfloor +1)}$, there are \textit{at most} $\lfloor \alpha \cardW \rfloor$ hypothetical assignments $\Wrep$ such that $\widehat{\tau}(\mathbf{Y}_\htau^{\imp},\Wrep)\geq\widehat{\tau}^\obs$. Thus, we have
$$
p(\htau)\leq \frac{\lfloor \alpha \cardW \rfloor}{\cardW}=\frac{ \alpha \cardW-( \alpha \cardW-\lfloor \alpha \cardW \rfloor)}{\cardW}\leq\frac{ \alpha \cardW}{\cardW}=\alpha \Rightarrow \htau\in\left\{\htau: p(\htau)\leq \alpha\right\},
$$
where the last inequality is due to $\alpha \cardW-\lfloor \alpha \cardW \rfloor\geq0$.

\end{proof}

\subsection{Proof of Theorem~\ref{thm:unbiased}}

\begin{proof}
We use $\Omega$ to denote the set of all chains $(w^{(0)},\ldots,w^{(T-1)},w^{(T)})$ that Algorithm \ref{alg:psrr} may generate. For any chain $(w^{(0)},\ldots,w^{(T-1)},w^{(T)})\in\Omega$ and $w=w^{(T)}$, we have

(a) 
$$
P\left(W^{(0)}=w^{(0)}\right)=P\left(W^{(0)}=1-w^{(0)}\right),
$$ 
since $W^{(0)}$ is generated from complete randomization;

(b) 
$$
P\left(W^{(t)}=w^{(t)}\mid W^{(t-1)}=w^{(t-1)}\right) = P\left(W^{(t)}=1-w^{(t)}\mid W^{(t-1)}=1-w^{(t-1)}\right),
$$ 
since $M(w^{(s)})=M(1-w^{(s)})$ for any $s$ and the transition probabilities only depend on the Mahalanobis distances with respect to two assignments. 

Therefore, we have
$$
\begin{aligned}
	&\pr\left(W^{(0)}=w^{(0)},\ldots,W^{(T-1)}=w^{(T-1)},W^{(T)}=w\right)\\
	=&\pr\left(W^{(0)}=w^{(0)}\right)\pr\left(W^{(1)}=w^{(1)}|W^{(0)}=w^{(0)}\right)\cdots \pr\left(W^{(T)}=w \mid W^{(T-1)}=w^{(T-1)}\right)  \\
	=&\pr\left(W^{(0)}=1-w^{(0)}\right)\pr\left(W^{(1)}=1-w^{(1)}|W^{(0)}=1-w^{(0)}\right)\cdots \\
	&\pr\left(W^{(T)}=1-w|W^{(T-1)}=1-w^{(T-1)}\right)  \\
	=&\pr\left(W^{(0)}=1-w^{(0)},\ldots,W^{(T-1)}=1-w^{(T-1)},W^{(T)}=1-w\right),
\end{aligned}
$$
which lead to $\pr(W=w) = \pr(W=1-w)$ by adding up the probabilities of the chains in $\Omega$. Thus,  $\pr(W_i=1)=\pr(W_i=0)=1/2$.
Then, 
$$
\begin{aligned}
	E\{\widehat\tau(W)\}&= E\left\{\frac2n\sum_{i=1}^n W_i Y_i(1) - \frac2n\sum_{i=1}^n (1-W_i)Y_i(0)\right\}\\
	&=\frac2n \sum_{i=1}^n Y_i(1)\pr(W_i=1) - \frac2n\sum_{i=1}^n Y_i(0)\pr(W_i=0)\\
	&=\tau.
\end{aligned}
$$
\end{proof}

\subsection{Proof of Theorem~\ref{thm:var}}

\begin{proof}

Let $\Sigma=\Cov_\cre\left(\overline X_t-\overline X_c\right)$ denote the covariance matrix of $\overline X_t-\overline X_c$ under complete randomization, and $Z=\Sigma^{-1/2}\left(\overline X_t-\overline X_c\right)$. Then the Mahalanobis distance can be expressed as 
$$
M=Z^\T Z=\sum_{j=1}^p Z_j^2.
$$

Then, we examine the covariance matrix of $Z$ under pair-switching rerandomization. For arguments similar to those in the proof of unbiasedness, we have $E_\psrr(Z_j)=0$. Thus, we have
$$
\Var_\psrr(Z_j) = E_\psrr(Z_j^2),\quad \Cov_\psrr(Z_j,Z_l) = E_\psrr(Z_j Z_l).
$$
Since $M\leq a$, we have
$$
E_\psrr\left(\sum_{j=1}^p Z_j^2\right)=E_\psrr(M)\leq a.
$$
If we exchange any $Z_j$ and $Z_l$, $M$ does not change. Thus, $Z_j$'s are exchangeable, thereby,
$$
v_a \equiv E_\psrr(Z_j^2)=E_\psrr\left(\sum_{j=1}^p Z_j^2\right)/p\leq a/p.
$$
If we change the sign of any $Z_j$, $M$ does not change. Thus, $(Z_j\mid Z_l)$ and $(-Z_j\mid Z_l)$ are identically distributed, thereby,
$$
\begin{aligned}
E_\psrr(Z_j Z_l)=&E_\psrr\{E_\psrr(Z_j Z_l\mid Z_l)\}\\
=&E_\psrr\{Z_j E_\psrr( Z_l\mid Z_l)\}\\
=&E_\psrr\{Z_j \times 0\}\\
=&0.
\end{aligned}
$$
Therefore, we have
$$
\begin{aligned}
\Cov_\psrr\left(\overline X_t-\overline X_c\right)=&\Sigma^{1/2}\Cov_\psrr(Z)(\Sigma^{1/2})^\T \\
=& \Sigma^{1/2}(v_a I)(\Sigma^{1/2})^\T \\
=&v_a\Sigma\\
=&v_a \Cov_\cre\left(\overline X_t-\overline X_c\right).
\end{aligned}
$$

Difference-in-means of outcomes can be expressed as 
$$
\widehat\tau=\tau +\beta^\T \left(\overline X_t-\overline X_c\right) + \left(\overline e_t-\overline e_c\right),
$$ 
where $\overline e_t=\sum_{i:W_i=1}e_{i}/n_t$ and $\overline e_c=\sum_{i:W_i=0}e_{i}/n_c$. Since $\beta_0+\beta^\T X_i$ is the projection of $Y_i(0)$ onto $(1, X)$, $\overline X_t-\overline X_c$ and $\overline e_t-\overline e_c$ are uncorrelated. Since $\widehat\tau$ and $\overline X_t-\overline X_c$ are normally distributed, $\overline X_t-\overline X_c$ and $\overline e_t-\overline e_c$ are independent.
Since pair-switching rerandomization does not affect $\overline e_t-\overline e_c$, we have
$$
\begin{aligned}
\Var_{\psrr}(\widehat\tau)=&\beta^\T\Cov_\psrr\left(\overline X_t-\overline X_c\right)\beta + \Var_\cre\left(\overline e_t-\overline e_c\right)\\
=&v_a \beta^\T\Cov_\cre\left(\overline X_t-\overline X_c\right)\beta + \Var_\cre\left(\overline e_t-\overline e_c\right) \\
=& v_a R^2 \Var_\cre(\widehat\tau) + (1-R^2)\Var_\cre(\widehat\tau)\\
=&\left\{1-(1-v_a)R^2\right\}\Var_\cre(\widehat\tau).
\end{aligned}
$$
Therefore, we have 
$$
\frac{\Var_{\cre}(\widehat\tau)-\Var_{\psrr}(\widehat\tau)}{\Var_{\cre}(\widehat\tau)}=\left(1-v_a\right)R^2\geq\left(1-a/p\right)R^2.
$$

\end{proof}

\subsection{Proof of Theorem~\ref{thm:seq_unbiased}}

\begin{proof}
When $K=1$, Algorithm \ref{alg:seqpsrr} is the same as Algorithm \ref{alg:psrr}.
Thus, according to the proof of Theorem \ref{thm:unbiased}, we have $\pr(W_i=1)=\pr(W_i=0)=1/2$ for $i\in [1]$, say, the units in the first group.

Then we consider the units in the $k$th group given the assignments of the first $k-1$ groups, $(W_{[1]},\ldots,W_{[k-1]}) = (w_{[1]},\ldots,w_{[k-1]})$. We use $\Omega_{[k]}$ to denote the set of all chains that Algorithm \ref{alg:seqpsrr} may generate in the $k$th group. For any chain $(w^{(0)}_{[k]},\ldots,w^{(T-1)}_{[k]},w^{(T)}_{[k]})\in\Omega$ and $w_{[k]}=w^{(T)}_{[k]}$, we have

(a) 
$$
P\left(W^{(0)}_{[k]}=w^{(0)}_{[k]}\right)=P\left(W^{(0)}_{[k]}=1-w^{(0)}_{[k]}\right)
$$ 
since $W^0_{[k]}$ is generated from complete randomization;

(b) 
$$
P\left(W^{(t)}_{[k]}=w^{(t)}_{[k]}\mid W^{(t-1)}_{[k]}=w^{(t-1)}_{[k]}\right) = P\left(W^{(t)}_{[k]}=1-w^{(t)}_{[k]}\mid W^{(t-1)}_{[k]}=1-w^{(t-1)}_{[k]}\right)
$$ 
since $M_k(w^{(s)}_{[k]})=M_k(1-w^{(s)}_{[k]})$ for any $s$ and the transition probabilities only depend on the Mahalanobis distances with respect to two assignments.

Therefore, we have
$$
\begin{aligned}
	&\pr\left(W^{(0)}_{[k]}=w^{(0)}_{[k]},\ldots,W^{(T-1)}_{[k]}=w^{(T-1)}_{[k]},W^{(T)}_{[k]}=w_{[k]}\right)\\
	=&\pr\left(W^{(0)}_{[k]}=w^{(0)}_{[k]}\right)\pr\left(W^{(1)}_{[k]}=w^{(1)}_{[k]}|W^{(0)}_{[k]}=w^{(0)}_{[k]}\right)\cdots \pr\left(W^{(T)}_{[k]}=w_{[k]}|W^{(T-1)}_{[k]}=w^{(T-1)}_{[k]}\right)  \\
	=&\pr\left(W^{(0)}_{[k]}=1-w^{(0)}_{[k]}\right)\pr\left(W^{(1)}_{[k]}=1-w^{(1)}_{[k]}|W^{(0)}_{[k]}=1-w^{(0)}_{[k]}\right)\cdots \\
	&\pr\left(W^{(T)}_{[k]}=1-w_{[k]}|W^{(T-1)}_{[k]}=1-w^{(T-1)}_{[k]}\right)  \\
	=&\pr\left(W^{(0)}_{[k]}=1-w^{(0)}_{[k]},\ldots,W^{(T-1)}_{[k]}=1-w^{(T-1)}_{[k]},W^{(T)}_{[k]}=1-w_{[k]}\right),
\end{aligned}
$$
which lead to $\pr(W_{[k]}=w_{[k]}) = \pr(W_{[k]}=1-w_{[k]})$ by adding up the probabilities of the paths in $\Omega_{[k]}$. Thus, $\pr(W_i=1)=\pr(W_i=0)=1/2$ for $i\in [k]$, say, the units in the $k$th group.

Considering all the groups together $W=(W_{[1]},\ldots,W_{[K]})$, we have
$$
\begin{aligned}
	E\{\widehat\tau(W)\}&= E\left\{\frac2n\sum_{i=1}^n W_iY_i(1) - \frac2n\sum_{i=1}^n (1-W_i)Y_i(0)\right\}\\
	&=\frac2n \sum_{i=1}^n Y_i(1)\pr(W_i=1) - \frac2n\sum_{i=1}^n Y_i(0)\pr(W_i=0)\\
	&=\tau.
\end{aligned}
$$
\end{proof}

\subsection{Proof of Theorem~\ref{thm:seq_var}}

\begin{proof}

The proof is essentially the same as the proof of Theorem \ref{thm:var}. Let $\Sigma=\Cov_\cre\left(\overline X_t-\overline X_c\right)$ denote the covariance matrix of $\overline X_t-\overline X_c$ under complete randomization, and $Z=\Sigma^{-1/2}\left(\overline X_t-\overline X_c\right)$. Then the Mahalanobis distance can be expressed as 
$$
M_K=Z^\T Z=\sum_{j=1}^p Z_j^2.
$$

Then, we examine the covariance matrix of $Z$ under sequential pair-switching rerandomization. For arguments similar to those in the proof of unbiasedness, we have $E_\seqpsrr(Z_j)=0$. Thus, we have
$$
\Var_\seqpsrr(Z_j) = E_\seqpsrr(Z_j^2),\quad \Cov_\seqpsrr(Z_j,Z_l) = E_\seqpsrr(Z_j Z_l).
$$
Since $M_K\leq a_K$, we have
$$
E_\seqpsrr\left(\sum_{j=1}^p Z_j^2\right)=E_\seqpsrr(M_K)\leq a_K.
$$
If we exchange any $Z_j$ and $Z_l$, $M_K$ does not change. Thus, $Z_j$'s are exchangeable, thereby,
$$
v_a \equiv E_\seqpsrr(Z_j^2)=E_\seqpsrr\left(\sum_{j=1}^p Z_j^2\right)/p\leq a_K/p.
$$
If we change the sign of any $Z_j$, $M_K$ does not change. Thus, $(Z_j\mid Z_l)$ and $(-Z_j\mid Z_l)$ are identically distributed, thereby,
$$
\begin{aligned}
E_\seqpsrr(Z_j Z_l)=&E_\seqpsrr\{E_\seqpsrr(Z_j Z_l\mid Z_l)\}\\
=&E_\seqpsrr\{Z_j E_\seqpsrr( Z_l\mid Z_l)\}\\
=&E_\seqpsrr\{Z_j \times 0\}\\
=&0.
\end{aligned}
$$
Therefore, we have
$$
\begin{aligned}
\Cov_\seqpsrr\left(\overline X_t-\overline X_c\right)=&\Sigma^{1/2}\Cov_\seqpsrr(Z)(\Sigma^{1/2})^\T \\
=& \Sigma^{1/2}(v_a I)(\Sigma^{1/2})^\T \\
=&v_a\Sigma\\
=&v_a \Cov_\cre\left(\overline X_t-\overline X_c\right).
\end{aligned}
$$

Difference-in-means of outcomes can be expressed as 
$$
\widehat\tau=\tau +\beta^\T \left(\overline X_t-\overline X_c\right) + \left(\overline e_t-\overline e_c\right),
$$ 
where $\overline e_t=\sum_{i:W_i=1}e_{i}/n_t$ and $\overline e_c=\sum_{i:W_i=0}e_{i}/n_c$. Since $\beta_0+\beta^\T X_i$ is the projection of $Y_i(0)$ onto $(1, X)$, $\overline X_t-\overline X_c$ and $\overline e_t-\overline e_c$ are uncorrelated. Since $\widehat\tau$ and $\overline X_t-\overline X_c$ are normally distributed, $\overline X_t-\overline X_c$ and $\overline e_t-\overline e_c$ are independent.
Since sequential pair-switching rerandomization does not affect $\overline e_t-\overline e_c$, we have
$$
\begin{aligned}
\Var_{\seqpsrr}(\widehat\tau)=&\beta^\T\Cov_\seqpsrr\left(\overline X_t-\overline X_c\right)\beta + \Var_\cre\left(\overline e_t-\overline e_c\right)\\
=&v_a \beta^\T\Cov_\cre\left(\overline X_t-\overline X_c\right)\beta + \Var_\cre\left(\overline e_t-\overline e_c\right) \\
=& v_a R^2 \Var_\cre(\widehat\tau) + (1-R^2)\Var_\cre(\widehat\tau)\\
=&\left\{1-(1-v_a)R^2\right\}\Var_\cre(\widehat\tau).
\end{aligned}
$$
Therefore, we have 
$$
\frac{\Var_{\cre}(\widehat\tau)-\Var_{\seqpsrr}(\widehat\tau)}{\Var_{\cre}(\widehat\tau)}=\left(1-v_a\right)R^2\geq\left(1-a_K/p\right)R^2.
$$

\end{proof}

\subsection{Calculation of Mahalanobis distance for pair-switching assignments}

\begin{proof}[Proof of Equation \eqref{eq:Mstar} in the main text]

Recall that $H=XS_{XX}^{-1}X^\T/\{n_t(1-n_t/n)\}$ and $h=(2n_t/n)H\mathbf{1}$ with $\mathbf{1}$ being an $n$-dimensional column vector of 1's. We further denote the propensity score by $p_w=n_t/n$. Then, we have
$$
\overline X_t-\overline X_c=\frac{X^{\T} W}{n_t}-\frac{X^{\T}(\mathbf{1}-W)}{n_c}=\frac{X^{\T}\left(W-p_{w} \mathbf{1}\right)}{n p_{w}\left(1-p_{w}\right)}.
$$
By the above equation and the definitions of $H$ and $h$, the Mahalanobis distance of any assignment $W$ can be expressed as
$$
\begin{aligned}
M(W)\equiv&\left(\overline X_t-\overline X_c\right)^{\T}\left[\Cov\left(\overline X_t-\overline X_c\right)\right]^{-1}\left(\overline X_t-\overline X_c\right)\\
=&n p_{w}\left(1-p_{w}\right)\left(\overline X_t-\overline X_c\right)^\T S_{XX}^{-1}\left(\overline X_t-\overline X_c\right)\\
=&\left(W-p_{w} \mathbf{1}\right)^{\T} \frac{X S_{XX}^{-1}X^{\T}}{n p_{w}\left(1-p_{w}\right)}\left(W-p_{w} \mathbf{1}\right)\\
=&\left(W-p_{w} \mathbf{1}\right)^{\T} H\left(W-p_{w} \mathbf{1}\right)\\
=&W^{\T} HW - 2p_wW^{\T} H\mathbf{1}+p_w^2 \mathbf{1}^\T H\mathbf{1} \\ 
=&W^{\T} HW - W^{\T} h+p_w^2\mathbf{1}^{\T}H\mathbf{1}.
\end{aligned}
$$
Thus, $W$ only affects the value of $M(W)$ through the first two terms.

We switch the $i$th and $j$th elements of $W^{(t)}$ with $W^{(t)}_i=1$ and $W^{(t)}_j=0$, and generate a new assignment $W^*$ with $W^*_i=0$ and $W^*_j=1$. The Mahalanobis distance of the assignment $W^{(t)}$ is denoted by $M^{(t)}=M(W^{(t)})$. We have
$$
\begin{aligned}
M(W^*)-M^{(t)}
=&\bigg\{(W^*)^{\T} HW^* - (W^*)^{\T} h\bigg\}-\bigg\{\left(W^{(t)}\right)^{\T} HW^{(t)} - \left(W^{(t)}\right)^{\T} h\bigg\}\\
=&\bigg\{(W^*)^{\T} HW^* - \left(W^{(t)}\right)^{\T} HW^{(t)} \bigg\}+ \bigg\{  \left(W^{(t)}\right)^{\T} h - (W^*)^{\T} h\bigg\}\\
=& \Big( 2\sum_{l\neq j } W^*_l H_{jl}+H_{jj} \Big) - \Big( 2\sum_{l\neq i} W^{(t)}_l H_{il}+H_{ii} \Big) + h_i - h_j\\
=& \Big( 2\sum_{l=1}^n W^*_l H_{jl}-H_{jj} \Big) - \Big( 2\sum_{l=1}^n W^{(t)}_l H_{il}-H_{ii} \Big) + h_i - h_j.
\end{aligned}
$$

\end{proof}

\section{Randomness metrics}
\label{sec:rand}

We denote the set of all pairs of $n$ units by $\R$ and the size of $\R$ is $|\R| = n!/\{(n-2)!2!\}=n(n-1)/2$. For any pair $\r\in\R$, under complete randomization, the probability of being in the same treatment groups is 
$$
\pcr\equiv 1- \left\{\frac{(n-2)!}{(n_t-1)!(n-n_t-1)!}\cdot \frac{2!}{1!1!}\right\} \left\{\frac{n!}{n_t!n_c!}\right\}^{-1}.
$$
When $n_t=n_c$, we have $\pcr=(n-2)/\{2(n-1)\}$. For a given randomization method, we can independently sample $B$ assignments from the distribution of $W$. Then, for every pair $\r\in\R$, we can compute the proportion of being in the same treatment groups in these $B$ assignments, denoted by $\{\prr:\r\in\R\}$. By comparing the difference between $\{\prr:\r\in\R\}$ and  $\pcr$, we can measure the deviation of the randomization procedure from complete randomization. 

The first randomness metric is defined as the normalized entropy of $\prr$,
$$
E_n=\frac{(|\R|)^{-1}\sum_{\r\in\R}\left\{\prr\log(\prr)+(1-\prr)\log(1-\prr)\right\}}{\pcr\log(\pcr)+(1-\pcr)\log(1-\pcr)},
$$
where $0\log(0)=0$. We have $0\leq E_n\leq 1$, $E_n=0$ if the design is deterministic, and $E_n=1$ under complete randomization.

The second randomness metric is defined as the normalized standard deviation of $\prr$,
$$
D_n = \left\{(|\R|)^{-1}\sum_{\r\in\R}(\prr-\pcr)^2\right\}^{1/2}\left\{(|\R|)^{-1}\sum_{\r\in\R_0}(0-\pcr)^2+ (|\R|)^{-1}\sum_{\r\in\R/\R_0}(1-\pcr)^2\right\}^{-1/2},
$$
where $\R_0$ is the set of pairs not in the same treatment groups for a single assignment, and $|\R_0|=n_tn_c$. Note that the mean of $\prr$ is equal to $\pcr$. We have $0\leq D_n\leq 1$, $D_n=1$ if the design is deterministic, and $D_n=0$ under complete randomization. \citet{krieger2019bka} only considered the case in which the sizes of treatment and control groups were equal. We generalize the definitions of $E_n$ and $D_n$ to the case of unequal treatment and control group sizes. Smaller values of $E_n$ and $D_n$ imply less randomness.

The last randomness metric is the maximum eigenvalue of the covariance matrix of $2W-1$, denoted by $L_n=\Lambda_{\max}(\Sigma_W)$. Under complete randomization, the covariance matrix $\Sigma_W$ is an identity matrix, thereby $L_n=1$. Larger value of $L_n$ implies less randomness.

Table \ref{tab:sim_rand} shows the Mahalanobis distance and three randomness metrics for different methods in our simulation studies.
In terms of three randomness metrics, both RR and PSRR are more random than GPS. Moreover, SeqPSRR is as random as SeqRR.

\section{Additional simulation results}
\label{sec:add_num}

\setcounter{table}{0}
\renewcommand{\thetable}{C\arabic{table}}
\setcounter{figure}{0}
\renewcommand{\thefigure}{C\arabic{figure}}

\subsection{Tuning parameter}

\begin{figure}[H]
\centering
    \includegraphics[scale = 0.85]{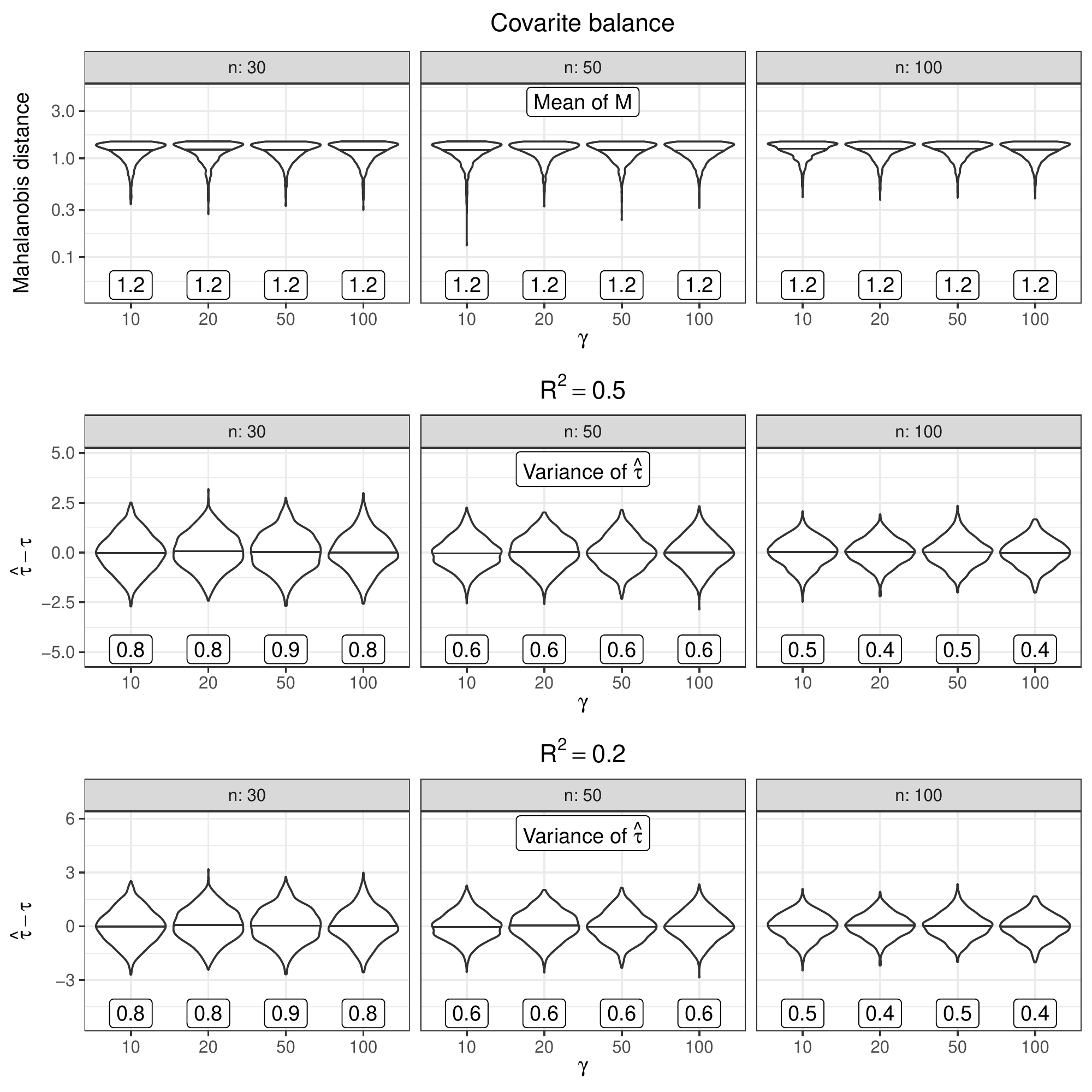}
\caption{Distributions of Mahalanobis distance and $\widehat\tau-\tau$ under PSRR with different values of the tuning parameter in the non-sequential setting.}
\label{fig:sim_tune_dist}
\end{figure}

\begin{figure}[H]
\centering
    \includegraphics[scale = 0.85]{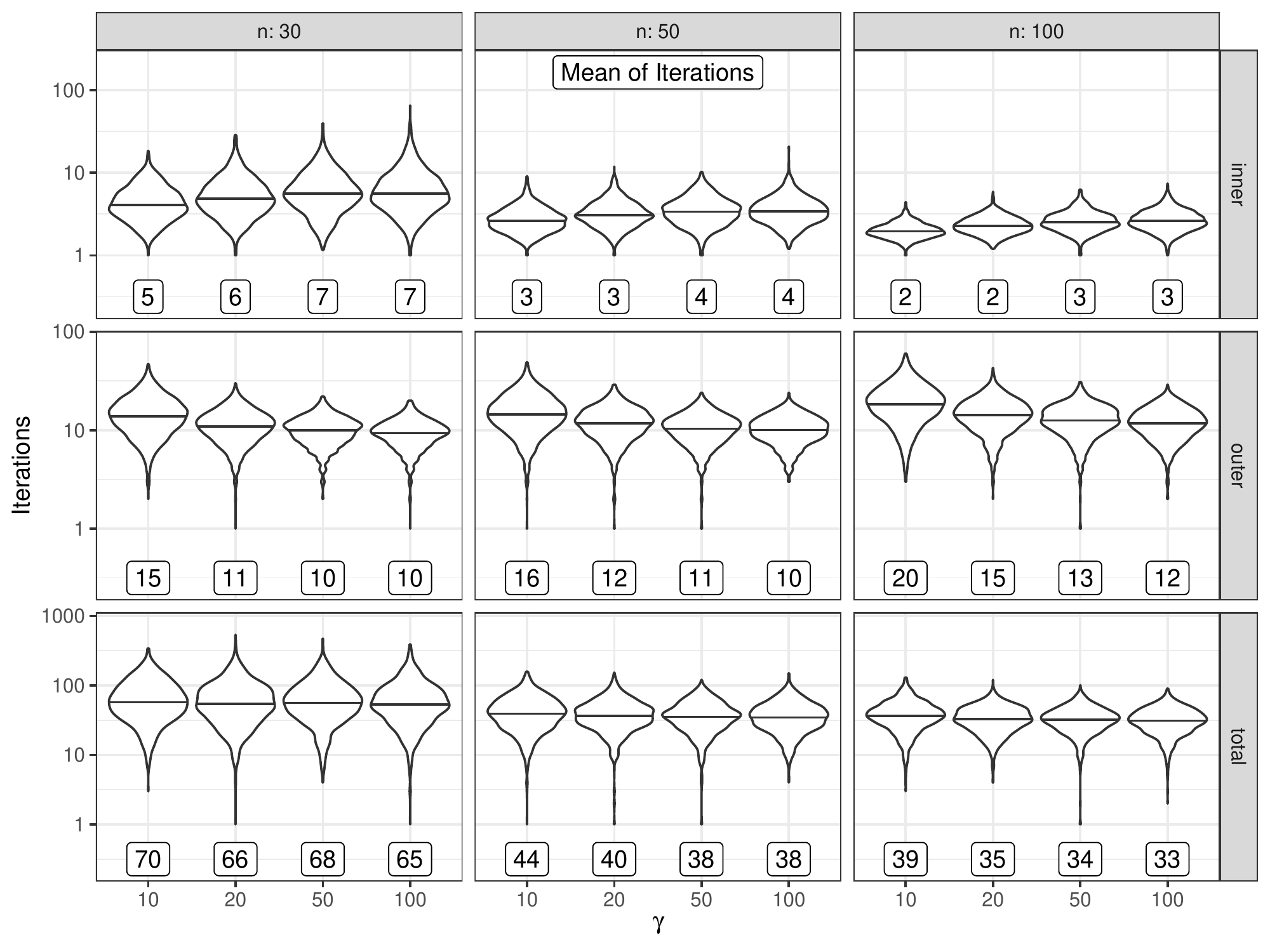}
\caption{Inner, outer, and total iterations of PSRR with different values of the tuning parameter to find one acceptable assignment in the non-sequential setting.}
\label{fig:sim_tune_iter}
\end{figure}

Tables \ref{tab:sim_tune_main}, \ref{tab:sim_tune_inf}, and Figure \ref{fig:sim_tune_dist} show that the statistical and computational performance of PSRR are similar using different values of the tuning parameter $\gamma$.
Figure \ref{fig:sim_tune_iter} shows the iterations of PSRR with different values of $\gamma$. As we expected, larger $\gamma$ leads to more iterations for PSRR to jump to a new candidate assignment, i.e., more inner iterations. However, the total iterations are similar for different values of $\gamma$.

\subsection{Small R-square}

\begin{figure}[H]
\centering
    \includegraphics[scale = 0.85]{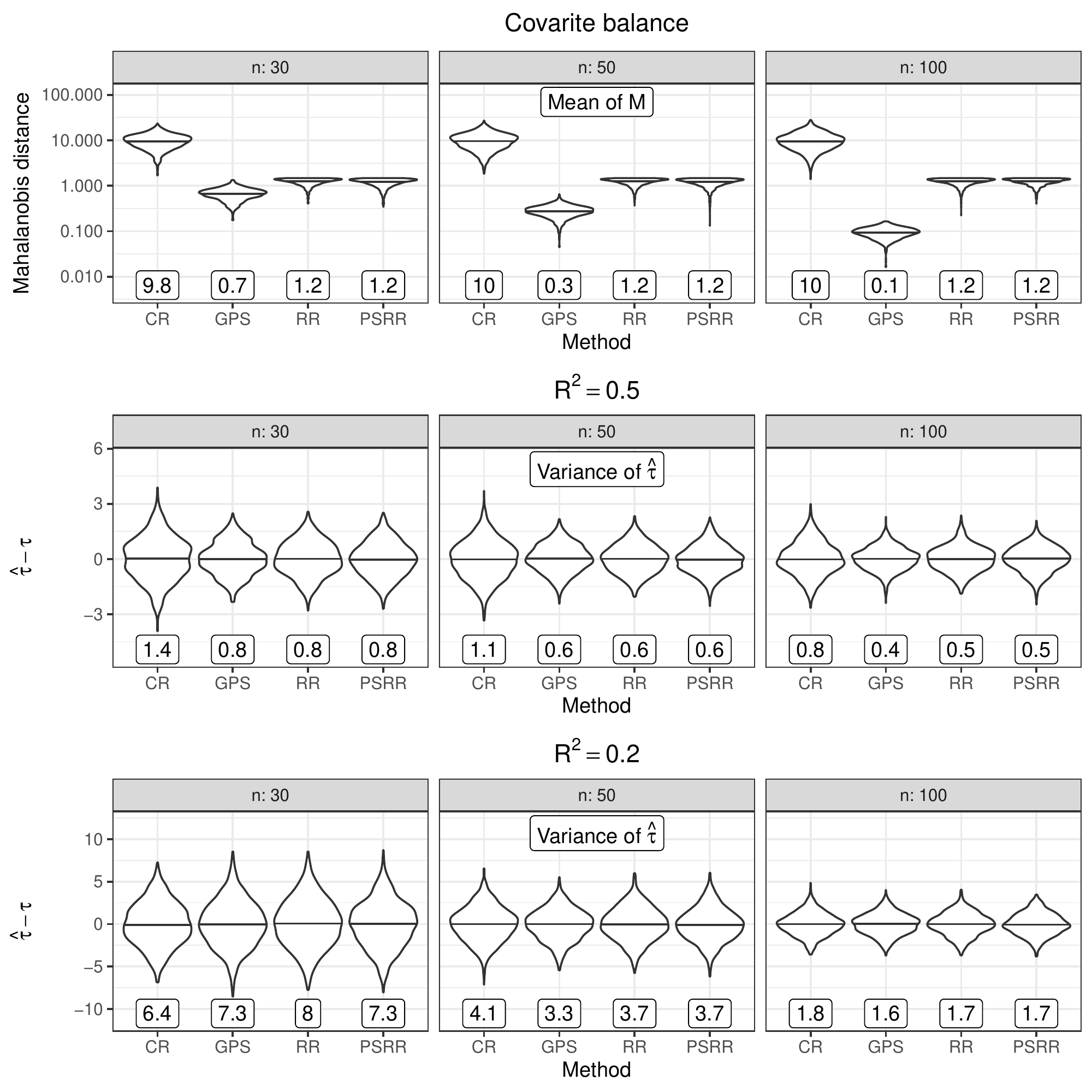}
\caption{Distributions of Mahalanobis distance and $\widehat\tau-\tau$ under different design methods in the non-sequential setting.}
\label{fig:sim_ns_dist}
\end{figure}

\begin{figure}[H]
\centering
    \includegraphics[scale = 0.85]{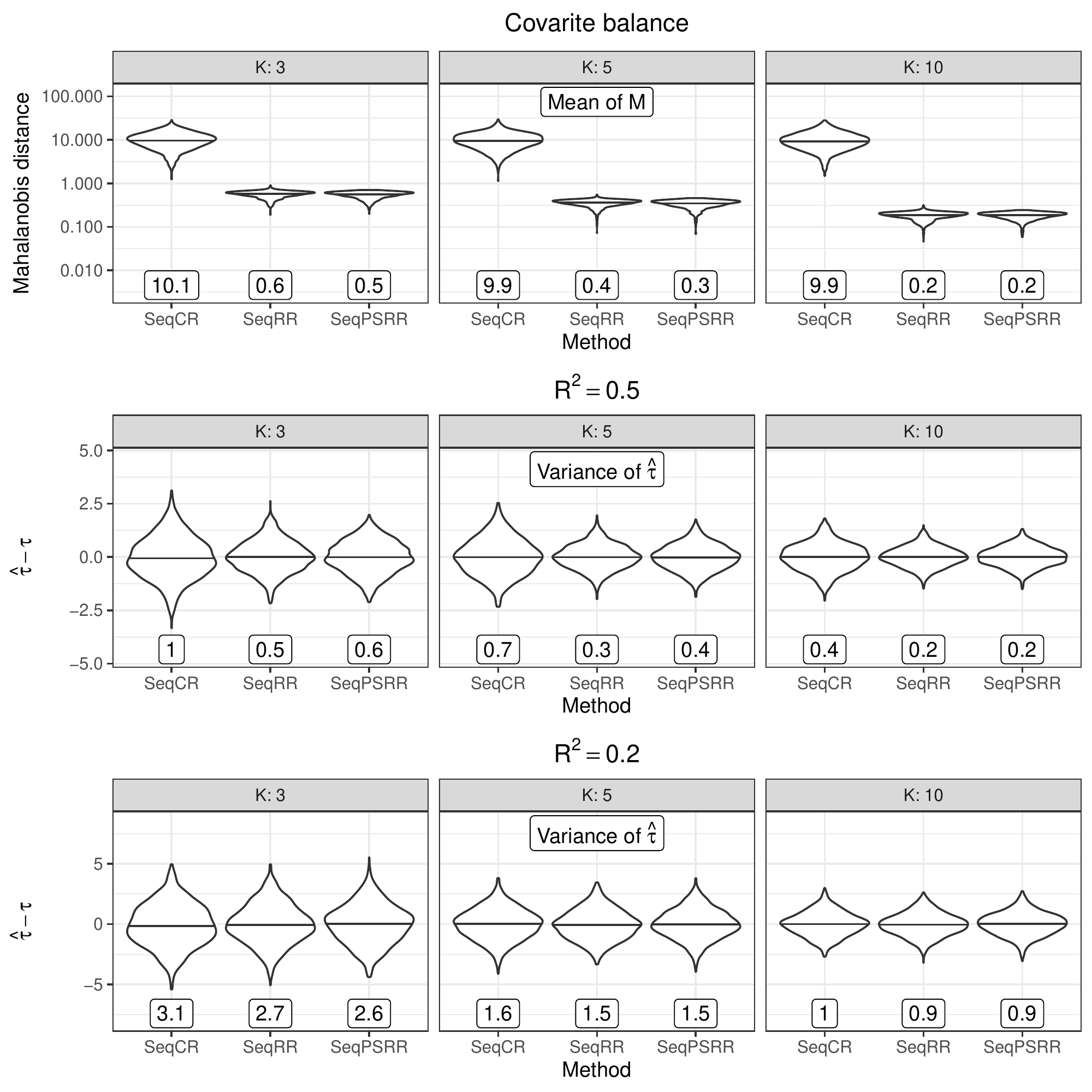}
\caption{Distributions of Mahalanobis distance and $\widehat\tau-\tau$ under different design methods in the sequential setting.}
\label{fig:sim_s_dist}
\end{figure}

Table \ref{tab:sim_ns_inf}, Figure \ref{fig:sim_ns_dist}, Table \ref{tab:sim_s_inf}, and Figure \ref{fig:sim_s_dist} show the additional simulation results for $R^2=0.2$ in the non-sequential setting and sequential setting, respectively.
Since the baseline covariates are the same for different $R^2$, we omit the results in the design stage and run times for $R^2=0.2$, which are almost the same as those for $R^2=0.5$. 
Under all designs, the biases of $\widehat{\tau}$ are also negligible, Fisher randomization tests control the type one error, and randomization-based confidence intervals reach the desired coverage rate.
Compared with CR/SeqCR, the improvements of other methods decrease when covariates are less predictive for the potential outcomes (i.e., $R^2$ is small). 

\subsection{Large sample size}

We conduct additional simulations with larger sample sizes. The simulation setup is the same as that in the main text, except that $n=500,1000,2000$ for non-sequentially randomized experiments and  $n_k= 50$ for sequentially randomized experiments. As the computation time of GPS increases rapidly with $n$, we do not include GPS in these simulations.
Table \ref{tab:sim_main_large_n} shows the results. We can see that PSRR and SeqPSRR are still applicable and the conclusions on computation time and efficiency are similar to those in Section \ref{sec:sim}.

\subsection{Constraint in the sequential setting}

\begin{figure}[H]
\centering
    \includegraphics[scale = 0.85]{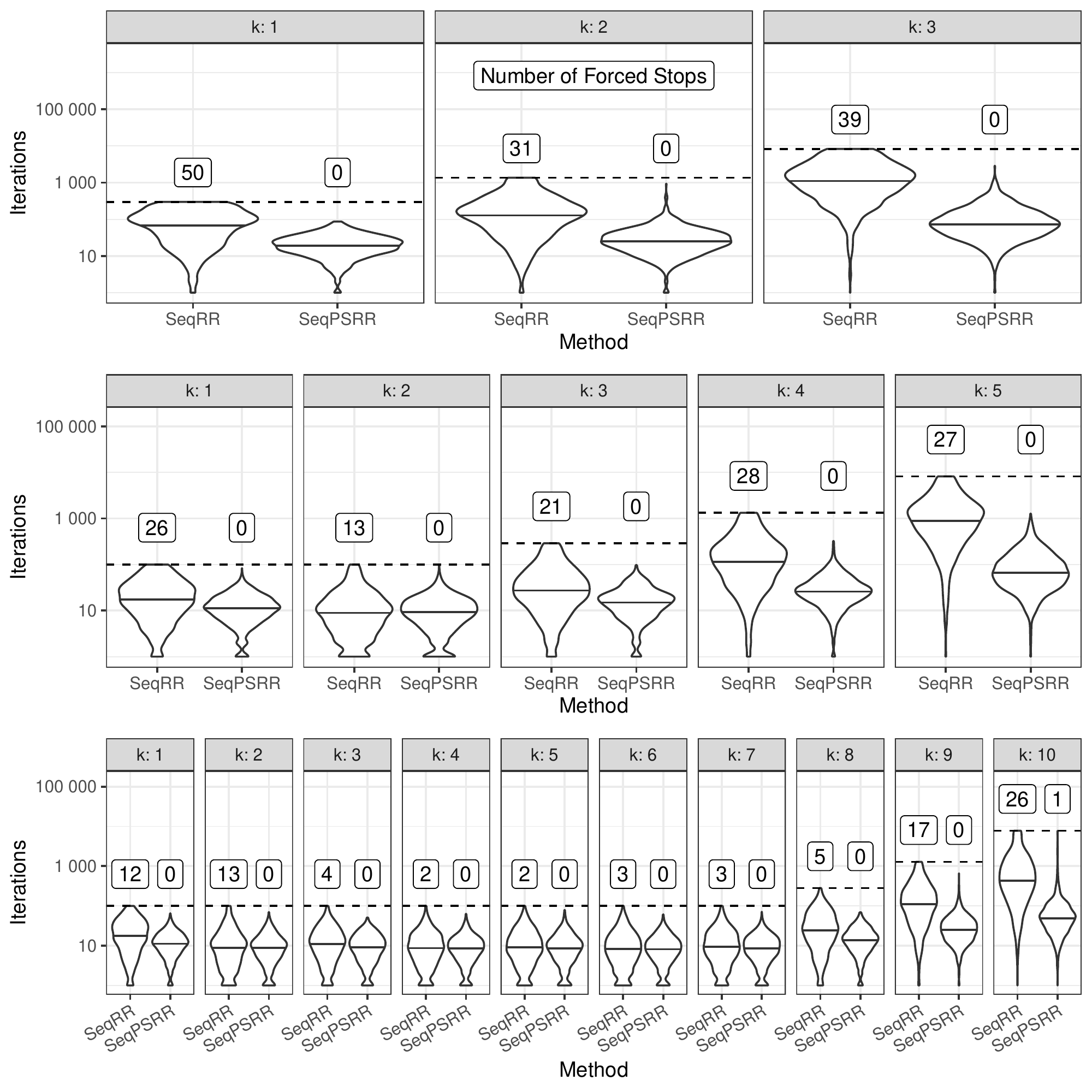}
\caption{Total iterations of SeqRR and SeqPSRR to find one acceptable assignment in each group in the sequential setting when $K=3,5,10$, respectively. The number in the rectangle is the frequency of forced stops (i.e., iterations reach $10s_k$) in 1000 replications.}
\label{fig:sim_s_iter}
\end{figure}

Figure \ref{fig:sim_s_iter} shows the total iterations of SeqRR and SeqPSRR under the $10s_k$ constraint.
First, this constraint does force SeqRR to stop iterating many times, preventing the number of total iterations of SeqRR from exceeding $10s_k$.
Second, this constraint does not have much effect on SeqPSRR because SeqPSRR is usually efficient enough to find the acceptable assignment before the iterations reach $10s_k$.
Third, when $K=10$ and $k=10$, we further examine the 27 cases where SeqRR or SeqPSRR is forced to stop. In each case, we fix the assignment for the first nine groups and enumerate all $20!/(10!10!)$ assignments for the units in the last group. In one case, there is no assignment such that $M_{[10]}\leq a_{10}$ at all, which means that an practical constraint on the total iteration is necessary to avoid infinite loop. Fortunately, SeqPSRR mitigated the impact of the constraint to some extent.

\subsection{Semi-synthetic datasets}

We first revisit the clinical trial introduced in Section \ref{sec:framework} and apply the proposed method on this dataset. Since we only observe half of the potential outcomes in the real dataset, for illustrative purpose, we impute the counterfactual potential outcomes under the sharp null.

We apply non-sequentially randomized experimental designs (CR, GPS, RR, and PSRR) to this dataset and perform subsequent estimation and inference in the same way as those in Section \ref{sec:sim_nonseq}. In this experiment, $n_t=20$ and $n_c=10$. We set $p_a=0.001$ for RR and PSRR. We replicate the estimation and inference procedure for $1000$ times to compare the repeated sampling properties of different methods.

\begin{figure}[H]
\centering
    \includegraphics[scale = 0.85]{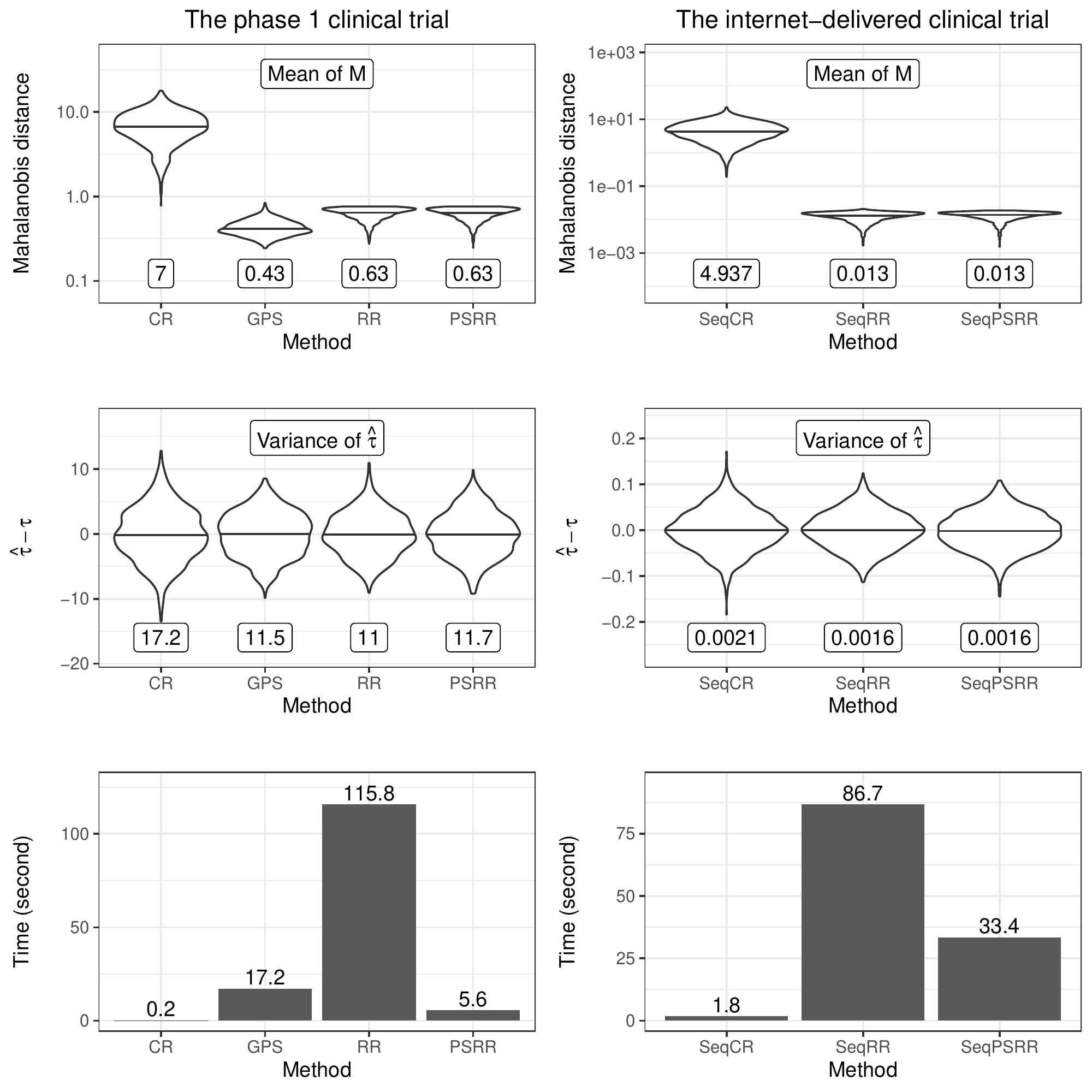}
\caption{Distributions of Mahalanobis distance ($M$), $\widehat\tau-\tau$ under different design methods, and run time in two clinical trial datasets.}
\label{fig:rd_dist}
\end{figure}

Table \ref{tab:rd_main} and the left panels of Figure \ref{fig:rd_dist} show the results. In the design stage, compared with CR, the other three designs reduce the imbalance of covariates. The distributions of Mahalanobis distance under RR and PSRR are similar. In the analysis stage, compared with CR, the other three methods (GPS, RR, and PSRR) reduce the standard deviation of difference-in-means estimator by 18\%--20\%, increase the power by 10\%--24\%, reduce the interval length by 19\%--20\%. Under all four designs, the distribution of $\widehat{\tau}-\tau$ are all symmetric at zero, confirming the unbiasedness of $\widehat{\tau}$. Moreover, PSRR is 3 times faster than GPS and 21 times faster than RR. Notably, compared with CR, the empirical PRIVs of PSRR is  32.0\%, which is close to the theoretical lower bound.

Then, we consider another clinical trial introduced in Section \ref{sec:seq}. The observed outcomes, abstinence at three-month follow-up, were binary and had missing values. For illustrative purpose, we first fit a logistic model to fill the missing outcomes, then impute the counterfactual outcomes under the sharp null.

We apply sequentially randomized experimental designs (SeqCR, SeqRR, and SeqPSRR) to this dataset and perform subsequent estimation and inference in the same way as those in Section \ref{sec:sim_seq}. We split $n=507$ participants into $K=15$ sequential groups (approximately waiting for one month for one group) with sizes $n_1=31$ and $n_k=34$, $k\geq 2$. We set $s_k=10$ for $k\leq 12$, $s_{13}=12$, $s_{14}=68$, and $s_{15}=800$. The experiments are replicated for $1000$ times to examine the repeated sampling properties. 

Table \ref{tab:rd_main} and the right panels of Figure \ref{fig:rd_dist} show the main results. In the design stage, compared with SeqCR, both SeqRR and SeqPSRR largely reduce the imbalance of covariates. The distributions of Mahalanobis distance under SeqRR and SeqPSRR are almost the same. In the analysis stage, compared with SeqCR, the other two methods (SeqRR and SeqPSRR) reduce the standard deviation of difference-in-means estimator by 12\%, increase the power by 13\%, reduce the interval length by 8\%. Under all three designs, the distribution of $\widehat{\tau}-\tau$ are all symmetric at zero, indicating the unbiasedness of $\widehat{\tau}$. Moreover, PSRR is approximately 2.6 times faster than SeqRR. Notably, compared with SeqCR, the empirical PRIVs of SeqPSRR is 22.4\%, which is larger than the theoretical lower bound ($13.8\%$).

\section*{Figures and Tables}


\begin{table}[H]

\caption{\label{tab:sim_rand}Statistical performance of different methods in the design stage.}
\centering
\begin{threeparttable}
\begin{tabular}[t]{l>{\centering\arraybackslash}p{2.5cm}>{\centering\arraybackslash}p{2.5cm}>{\centering\arraybackslash}p{2.5cm}>{\centering\arraybackslash}p{2.5cm}}
\toprule
\multicolumn{1}{c}{Method} & \multicolumn{1}{c}{$M$} & \multicolumn{1}{c}{$E_n$} & \multicolumn{1}{c}{$D_n$} & \multicolumn{1}{c}{$L_n$}\\
\midrule
\addlinespace[0.3em]
\multicolumn{5}{c}{\textbf{Non-sequential: $n=30$}}\\
\hspace{1em}CR & 9.789 & 0.999 & 0.032 & 1.371\\
\hspace{1em}GPS & 0.667 & 0.985 & 0.143 & 2.476\\
\hspace{1em}RR & 1.224 & 0.987 & 0.134 & 2.434\\
\hspace{1em}PSRR & 1.180 & 0.987 & 0.134 & 2.365\\
\midrule
\addlinespace[0.3em]
\multicolumn{5}{c}{\textbf{Non-sequential: $n=50$}}\\
\hspace{1em}CR & 9.969 & 0.999 & 0.032 & 1.483\\
\hspace{1em}GPS & 0.279 & 0.995 & 0.080 & 1.904\\
\hspace{1em}RR & 1.222 & 0.996 & 0.073 & 1.846\\
\hspace{1em}PSRR & 1.180 & 0.996 & 0.073 & 1.826\\
\midrule
\addlinespace[0.3em]
\multicolumn{5}{c}{\textbf{Non-sequential: $n=100$}}\\
\hspace{1em}CR & 9.997 & 0.999 & 0.032 & 1.707\\
\hspace{1em}GPS & 0.094 & 0.998 & 0.046 & 1.870\\
\hspace{1em}RR & 1.217 & 0.999 & 0.044 & 1.842\\
\hspace{1em}PSRR & 1.225 & 0.999 & 0.044 & 1.840\\
\midrule
\addlinespace[0.3em]
\multicolumn{5}{c}{\textbf{Sequential: $K=3, n_k=20$}}\\
\hspace{1em}SeqCR & 10.069 & 0.999 & 0.040 & 1.573\\
\hspace{1em}SeqRR & 0.559 & 0.992 & 0.105 & 3.345\\
\hspace{1em}SeqPSRR & 0.542 & 0.992 & 0.105 & 3.138\\
\midrule
\addlinespace[0.3em]
\multicolumn{5}{c}{\textbf{Sequential: $K=5, n_k=20$}}\\
\hspace{1em}SeqCR & 9.913 & 0.999 & 0.038 & 1.762\\
\hspace{1em}SeqRR & 0.353 & 0.997 & 0.069 & 2.770\\
\hspace{1em}SeqPSRR & 0.337 & 0.997 & 0.069 & 2.667\\
\midrule
\addlinespace[0.3em]
\multicolumn{5}{c}{\textbf{Sequential: $K=10, n_k=20$}}\\
\hspace{1em}SeqCR & 9.929 & 0.999 & 0.035 & 2.133\\
\hspace{1em}SeqRR & 0.181 & 0.998 & 0.048 & 2.959\\
\hspace{1em}SeqPSRR & 0.180 & 0.998 & 0.049 & 2.867\\
\bottomrule
\end{tabular}
\begin{tablenotes}
\item Note: $M$, mean Mahalanobis distance; $E_n$, $D_n$, and $L_n$, randomness metrics.
\end{tablenotes}
\end{threeparttable}
\end{table}

\begin{table}[H]

\caption{\label{tab:sim_tune_main}Statistical and computational performance of PSRR with different values of the tuning parameter in the non-sequential setting when $R^2=0.5$.}
\centering
\begin{threeparttable}
\begin{tabular}[t]{rrrrrrrrrrrrrr}
\toprule
\multicolumn{1}{c}{ } & \multicolumn{4}{c}{Design ($\times 10^{-2}$)} & \multicolumn{6}{c}{Inference ($\times 10^{-2}$)} & \multicolumn{3}{c}{Run time (sec.)} \\
\cmidrule(l{3pt}r{3pt}){2-5} \cmidrule(l{3pt}r{3pt}){6-11} \cmidrule(l{3pt}r{3pt}){12-14}
\multicolumn{1}{c}{$\gamma$} & \multicolumn{1}{c}{$M$} & \multicolumn{1}{c}{$E_n$} & \multicolumn{1}{c}{$D_n$} & \multicolumn{1}{c}{$L_n$} & \multicolumn{1}{c}{Bias} & \multicolumn{1}{c}{SD} & \multicolumn{1}{c}{Size} & \multicolumn{1}{c}{Pow.} & \multicolumn{1}{c}{CP} & \multicolumn{1}{c}{Len.} & \multicolumn{1}{c}{Sample} & \multicolumn{1}{c}{Ex.} & \multicolumn{1}{c}{Bi.}\\
\midrule
\addlinespace[0.3em]
\multicolumn{14}{c}{\textbf{$n=30$}}\\
\hspace{1em}10 & 118 & 99 & 13 & 237 & 2.2 & 92 & 4.2 & 15 & 95.8 & 392 & 5.2 & 0.02 & 0.61\\
\hspace{1em}20 & 119 & 99 & 13 & 235 & 7.7 & 90 & 4.2 & 17 & 95.8 & 392 & 4.9 & 0.02 & 0.62\\
\hspace{1em}50 & 119 & 99 & 13 & 233 & 3.8 & 93 & 5.3 & 17 & 94.7 & 391 & 5.0 & 0.02 & 0.62\\
\hspace{1em}100 & 118 & 99 & 13 & 233 & 2.4 & 92 & 5.1 & 15 & 94.9 & 392 & 5.3 & 0.02 & 0.61\\
\midrule
\addlinespace[0.3em]
\multicolumn{14}{c}{\textbf{$n=50$}}\\
\hspace{1em}10 & 118 & 100 & 7 & 183 & 2.1 & 75 & 4.0 & 24 & 96.0 & 319 & 3.6 & 0.02 & 0.57\\
\hspace{1em}20 & 120 & 100 & 7 & 182 & 3.4 & 80 & 5.1 & 30 & 94.9 & 319 & 3.3 & 0.02 & 0.57\\
\hspace{1em}50 & 118 & 100 & 7 & 182 & 1.8 & 78 & 4.3 & 24 & 95.7 & 319 & 3.0 & 0.02 & 0.58\\
\hspace{1em}100 & 118 & 100 & 7 & 182 & 0.6 & 79 & 5.0 & 27 & 95.0 & 319 & 3.0 & 0.02 & 0.58\\
\midrule
\addlinespace[0.3em]
\multicolumn{14}{c}{\textbf{$n=100$}}\\
\hspace{1em}10 & 122 & 100 & 4 & 184 & 0.2 & 69 & 4.9 & 50 & 95.1 & 265 & 3.2 & 0.03 & 0.60\\
\hspace{1em}20 & 122 & 100 & 4 & 184 & 3.0 & 63 & 3.8 & 50 & 96.2 & 265 & 2.8 & 0.02 & 0.60\\
\hspace{1em}50 & 122 & 100 & 4 & 184 & 2.2 & 69 & 5.7 & 50 & 94.3 & 265 & 2.7 & 0.02 & 0.59\\
\hspace{1em}100 & 120 & 100 & 4 & 184 & 2.7 & 67 & 5.7 & 48 & 94.3 & 265 & 2.6 & 0.03 & 0.62\\
\bottomrule
\end{tabular}
\begin{tablenotes}
\item Note: $M$, mean Mahalanobis distance; $E_n$, $D_n$, and $L_n$, randomness metrics; Bias, absolute bias; SD, standard deviation; Pow., power; CP, coverage probability; Len., mean interval length; Sample, sampling $10^3$ acceptable assignments; Ex./Bi., determining interval endpoints by the exact approach/bisection method.
\end{tablenotes}
\end{threeparttable}
\end{table}

\begin{table}[H]

\caption{\label{tab:sim_tune_inf}Inference results of PSRR with different values of the tuning parameter in the non-sequential setting when $R^2=0.2$.}
\centering
\begin{threeparttable}
\begin{tabular}[t]{r>{\raggedleft\arraybackslash}p{1.6cm}>{\raggedleft\arraybackslash}p{1.6cm}>{\raggedleft\arraybackslash}p{1.6cm}>{\raggedleft\arraybackslash}p{1.6cm}>{\raggedleft\arraybackslash}p{1.6cm}>{\raggedleft\arraybackslash}p{1.6cm}}
\toprule
\multicolumn{1}{r}{$\gamma$} & \multicolumn{1}{>{\raggedleft\arraybackslash}p{1.6cm}}{Bias} & \multicolumn{1}{>{\raggedleft\arraybackslash}p{1.6cm}}{SD} & \multicolumn{1}{>{\raggedleft\arraybackslash}p{1.6cm}}{Size} & \multicolumn{1}{>{\raggedleft\arraybackslash}p{1.6cm}}{Pow.} & \multicolumn{1}{>{\raggedleft\arraybackslash}p{1.6cm}}{CP} & \multicolumn{1}{>{\raggedleft\arraybackslash}p{1.6cm}}{Len.}\\
\midrule
\addlinespace[0.3em]
\multicolumn{7}{c}{\textbf{$n=30$}}\\
\hspace{1em}10 & 2.7 & 270 & 5.2 & 10 & 94.8 & 1157\\
\hspace{1em}20 & 1.2 & 274 & 5.7 & 10 & 94.3 & 1154\\
\hspace{1em}50 & 4.8 & 258 & 2.8 & 9 & 97.2 & 1159\\
\hspace{1em}100 & 6.9 & 272 & 5.3 & 10 & 94.7 & 1154\\
\midrule
\addlinespace[0.3em]
\multicolumn{7}{c}{\textbf{$n=50$}}\\
\hspace{1em}10 & 8.3 & 193 & 6.0 & 17 & 94.0 & 775\\
\hspace{1em}20 & 14.9 & 198 & 6.2 & 22 & 93.8 & 775\\
\hspace{1em}50 & 7.0 & 197 & 5.7 & 22 & 94.3 & 774\\
\hspace{1em}100 & 3.6 & 189 & 4.3 & 20 & 95.7 & 776\\
\midrule
\addlinespace[0.3em]
\multicolumn{7}{c}{\textbf{$n=100$}}\\
\hspace{1em}10 & 3.2 & 129 & 5.8 & 33 & 94.2 & 506\\
\hspace{1em}20 & 0.9 & 128 & 4.6 & 35 & 95.4 & 507\\
\hspace{1em}50 & 3.2 & 125 & 5.4 & 32 & 94.6 & 507\\
\hspace{1em}100 & 6.7 & 130 & 6.5 & 31 & 93.5 & 507\\
\bottomrule
\end{tabular}
\begin{tablenotes}
\item Note: Bias, absolute bias; SD, standard deviation; Pow., power; CP, coverage probability; Len., mean interval length. The values in the last six columns are multiplied by 100.
\end{tablenotes}
\end{threeparttable}
\end{table}


\begin{table}[H]

\caption{\label{tab:sim_ns_inf}Inference results of different methods in the non-sequential setting when $R^2=0.2$.}
\centering
\begin{threeparttable}
\begin{tabular}[t]{l>{\raggedleft\arraybackslash}p{1.6cm}>{\raggedleft\arraybackslash}p{1.6cm}>{\raggedleft\arraybackslash}p{1.6cm}>{\raggedleft\arraybackslash}p{1.6cm}>{\raggedleft\arraybackslash}p{1.6cm}>{\raggedleft\arraybackslash}p{1.6cm}}
\toprule
\multicolumn{1}{r}{Method} & \multicolumn{1}{>{\raggedleft\arraybackslash}p{1.6cm}}{Bias} & \multicolumn{1}{>{\raggedleft\arraybackslash}p{1.6cm}}{SD} & \multicolumn{1}{>{\raggedleft\arraybackslash}p{1.6cm}}{Size} & \multicolumn{1}{>{\raggedleft\arraybackslash}p{1.6cm}}{Pow.} & \multicolumn{1}{>{\raggedleft\arraybackslash}p{1.6cm}}{CP} & \multicolumn{1}{>{\raggedleft\arraybackslash}p{1.6cm}}{Len.}\\
\midrule
\addlinespace[0.3em]
\multicolumn{7}{c}{\textbf{$n=30$}}\\
\hspace{1em}CR & 3.4 & 253 & 3.6 & 10 & 96.4 & 1077\\
\hspace{1em}GPS & 1.6 & 271 & 5.0 & 10 & 95.0 & 1172\\
\hspace{1em}RR & 9.8 & 283 & 5.7 & 12 & 94.3 & 1154\\
\hspace{1em}PSRR & 2.7 & 270 & 5.2 & 10 & 94.8 & 1157\\
\midrule
\addlinespace[0.3em]
\multicolumn{7}{c}{\textbf{$n=50$}}\\
\hspace{1em}CR & 2.0 & 203 & 4.2 & 16 & 95.8 & 837\\
\hspace{1em}GPS & 6.0 & 180 & 3.9 & 18 & 96.1 & 769\\
\hspace{1em}RR & 0.3 & 194 & 5.2 & 20 & 94.8 & 774\\
\hspace{1em}PSRR & 8.3 & 193 & 6.0 & 17 & 94.0 & 775\\
\midrule
\addlinespace[0.3em]
\multicolumn{7}{c}{\textbf{$n=100$}}\\
\hspace{1em}CR & 0.5 & 132 & 6.0 & 28 & 94.0 & 527\\
\hspace{1em}GPS & 0.7 & 125 & 5.5 & 34 & 94.5 & 501\\
\hspace{1em}RR & 1.5 & 131 & 6.0 & 33 & 94.0 & 505\\
\hspace{1em}PSRR & 3.2 & 129 & 5.8 & 33 & 94.2 & 506\\
\bottomrule
\end{tabular}
\begin{tablenotes}
\item Note: Bias, absolute bias; SD, standard deviation; Pow., power; CP, coverage probability; Len., mean interval length. The values in the last six columns are multiplied by 100.
\end{tablenotes}
\end{threeparttable}
\end{table}


\begin{table}[H]

\caption{\label{tab:sim_s_inf}Inference results of different methods in the sequential setting when $R^2=0.2$.}
\centering
\begin{threeparttable}
\begin{tabular}[t]{l>{\raggedleft\arraybackslash}p{1.6cm}>{\raggedleft\arraybackslash}p{1.6cm}>{\raggedleft\arraybackslash}p{1.6cm}>{\raggedleft\arraybackslash}p{1.6cm}>{\raggedleft\arraybackslash}p{1.6cm}>{\raggedleft\arraybackslash}p{1.6cm}}
\toprule
\multicolumn{1}{r}{Method} & \multicolumn{1}{>{\raggedleft\arraybackslash}p{1.6cm}}{Bias} & \multicolumn{1}{>{\raggedleft\arraybackslash}p{1.6cm}}{SD} & \multicolumn{1}{>{\raggedleft\arraybackslash}p{1.6cm}}{Size} & \multicolumn{1}{>{\raggedleft\arraybackslash}p{1.6cm}}{Pow.} & \multicolumn{1}{>{\raggedleft\arraybackslash}p{1.6cm}}{CP} & \multicolumn{1}{>{\raggedleft\arraybackslash}p{1.6cm}}{Len.}\\
\midrule
\addlinespace[0.3em]
\multicolumn{7}{c}{\textbf{$K=3, n_k=20$}}\\
\hspace{1em}SeqCR & 11.0 & 176 & 4.7 & 18 & 95.3 & 715\\
\hspace{1em}SeqRR & 0.6 & 163 & 5.3 & 22 & 94.7 & 665\\
\hspace{1em}SeqPSRR & 0.5 & 162 & 4.5 & 21 & 95.5 & 667\\
\midrule
\addlinespace[0.3em]
\multicolumn{7}{c}{\textbf{$K=5, n_k=20$}}\\
\hspace{1em}SeqCR & 0.1 & 127 & 4.4 & 32 & 95.6 & 523\\
\hspace{1em}SeqRR & 4.9 & 123 & 4.9 & 32 & 95.1 & 495\\
\hspace{1em}SeqPSRR & 0.1 & 122 & 4.7 & 33 & 95.3 & 498\\
\midrule
\addlinespace[0.3em]
\multicolumn{7}{c}{\textbf{$K=10, n_k=20$}}\\
\hspace{1em}SeqCR & 0.5 & 98 & 4.3 & 56 & 95.7 & 385\\
\hspace{1em}SeqRR & 3.6 & 93 & 5.6 & 59 & 94.4 & 362\\
\hspace{1em}SeqPSRR & 1.2 & 93 & 5.3 & 61 & 94.7 & 364\\
\bottomrule
\end{tabular}
\begin{tablenotes}
\item Note: Bias, absolute bias; SD, standard deviation; Pow., power; CP, coverage probability; Len., mean interval length. The values in the last six columns are multiplied by 100.
\end{tablenotes}
\end{threeparttable}
\end{table}


\begin{table}[H]

\caption{\label{tab:sim_main_large_n}Statistical and computational performance of different methods for large sample size.}
\centering
\begin{threeparttable}
\begin{tabular}[t]{l>{\raggedleft\arraybackslash}p{1.1cm}>{\raggedleft\arraybackslash}p{1.1cm}>{\raggedleft\arraybackslash}p{1.1cm}>{\raggedleft\arraybackslash}p{1.1cm}>{\raggedleft\arraybackslash}p{1.1cm}>{\raggedleft\arraybackslash}p{1.1cm}>{\raggedleft\arraybackslash}p{1.1cm}>{\raggedleft\arraybackslash}p{1.1cm}>{\raggedleft\arraybackslash}p{1.1cm}}
\toprule
\multicolumn{1}{c}{ } & \multicolumn{6}{c}{Inference ($\times 10^{-2}$)} & \multicolumn{3}{c}{Run time (sec.)} \\
\cmidrule(l{3pt}r{3pt}){2-7} \cmidrule(l{3pt}r{3pt}){8-10}
\multicolumn{1}{c}{Method} & \multicolumn{1}{>{\centering\arraybackslash}p{1.1cm}}{Bias} & \multicolumn{1}{>{\centering\arraybackslash}p{1.1cm}}{SD} & \multicolumn{1}{>{\centering\arraybackslash}p{1.1cm}}{Size} & \multicolumn{1}{>{\centering\arraybackslash}p{1.1cm}}{Pow.} & \multicolumn{1}{>{\centering\arraybackslash}p{1.1cm}}{CP} & \multicolumn{1}{>{\centering\arraybackslash}p{1.1cm}}{Len.} & \multicolumn{1}{>{\centering\arraybackslash}p{1.1cm}}{Sample} & \multicolumn{1}{>{\centering\arraybackslash}p{1.1cm}}{Ex.} & \multicolumn{1}{>{\centering\arraybackslash}p{1.1cm}}{Bi.}\\
\midrule
\addlinespace[0.3em]
\multicolumn{10}{c}{\textbf{Non-sequential: $n=500$}}\\
\hspace{1em}CR & 0.3 & 43 & 5.8 & 91 & 94.2 & 164 & 0.3 & 0.04 & 1.06\\
\hspace{1em}RR & 1.3 & 32 & 4.8 & 100 & 95.2 & 122 & 99.7 & 0.04 & 1.04\\
\hspace{1em}PSRR & 0.5 & 31 & 4.6 & 100 & 95.4 & 123 & 9.1 & 0.04 & 0.95\\
\midrule
\addlinespace[0.3em]
\multicolumn{10}{c}{\textbf{Non-sequential: $n=1000$}}\\
\hspace{1em}CR & 0.2 & 28 & 4.2 & 100 & 95.8 & 113 & 0.3 & 0.07 & 1.32\\
\hspace{1em}RR & 0.7 & 22 & 5.5 & 100 & 94.5 & 86 & 146.9 & 0.07 & 1.30\\
\hspace{1em}PSRR & 0.5 & 21 & 4.3 & 100 & 95.7 & 86 & 19.6 & 0.07 & 1.34\\
\midrule
\addlinespace[0.3em]
\multicolumn{10}{c}{\textbf{Non-sequential: $n=2000$}}\\
\hspace{1em}CR & 0.1 & 20 & 5.0 & 100 & 95.0 & 78 & 0.4 & 0.12 & 1.93\\
\hspace{1em}RR & 0.5 & 14 & 3.8 & 100 & 96.2 & 59 & 259.4 & 0.14 & 2.23\\
\hspace{1em}PSRR & 0.4 & 15 & 4.5 & 100 & 95.5 & 59 & 48.7 & 0.14 & 2.21\\
\midrule
\addlinespace[0.3em]
\multicolumn{10}{c}{\textbf{Sequential: $K=3, n_k=50$}}\\
\hspace{1em}SeqCR & 2.8 & 78 & 5.8 & 46 & 94.2 & 305 & 0.4 & 0.02 & 0.63\\
\hspace{1em}SeqRR & 3.0 & 52 & 4.7 & 77 & 95.3 & 206 & 58.0 & 0.02 & 0.63\\
\hspace{1em}SeqPSRR & 2.2 & 52 & 5.1 & 77 & 94.9 & 207 & 9.4 & 0.02 & 0.65\\
\midrule
\addlinespace[0.3em]
\multicolumn{10}{c}{\textbf{Sequential: $K=5, n_k=50$}}\\
\hspace{1em}SeqCR & 1.1 & 59 & 5.0 & 65 & 95.0 & 230 & 0.7 & 0.02 & 0.80\\
\hspace{1em}SeqRR & 0.9 & 39 & 4.7 & 94 & 95.3 & 159 & 85.7 & 0.03 & 0.76\\
\hspace{1em}SeqPSRR & 1.3 & 39 & 3.9 & 93 & 96.1 & 159 & 14.2 & 0.02 & 0.69\\
\midrule
\addlinespace[0.3em]
\multicolumn{10}{c}{\textbf{Sequential: $K=10, n_k=50$}}\\
\hspace{1em}SeqCR & 0.4 & 42 & 5.1 & 90 & 94.9 & 164 & 1.6 & 0.04 & 0.95\\
\hspace{1em}SeqRR & 0.0 & 29 & 5.1 & 99 & 94.9 & 114 & 114.5 & 0.04 & 1.04\\
\hspace{1em}SeqPSRR & 3.2 & 29 & 5.9 & 100 & 94.1 & 114 & 24.7 & 0.04 & 0.92\\
\bottomrule
\end{tabular}
\begin{tablenotes}
\item Note: $M$, mean Mahalanobis distance; $E_n$, $D_n$, and $L_n$, randomness metrics; Bias, absolute bias; SD, standard deviation; Pow., power; CP, coverage probability; Len., mean interval length; Sample, sampling $10^3$ acceptable assignments; Ex./Bi., determining interval endpoints by the exact approach/bisection method.
\end{tablenotes}
\end{threeparttable}
\end{table}


\begin{table}[H]

\caption{\label{tab:rd_reserpine_dif}Summary statistics for the reserpine dataset.}
\centering
\begin{threeparttable}
\begin{tabular}[t]{lrr>{\raggedleft\arraybackslash}p{2.5cm}r}
\toprule
\multicolumn{1}{c}{Variable} & \multicolumn{1}{c}{Reserpine} & \multicolumn{1}{c}{Placebo} & \multicolumn{1}{>{\centering\arraybackslash}p{2.5cm}}{Standardized difference} & \multicolumn{1}{c}{$p$-value}\\
\midrule
\addlinespace[0.3em]
\multicolumn{5}{l}{\textbf{Baseline}}\\
\hspace{1em}Gender (percentage of males) & 85 & 80 & 0.13 & 0.75\\
\hspace{1em}Age & 28 & 32 & -0.52 & 0.21\\
\hspace{1em}Height & 176 & 179 & -0.40 & 0.32\\
\hspace{1em}Weight & 156 & 174 & -0.83 & 0.05\\
\hspace{1em}Systolic blood pressure & 121 & 118 & 0.23 & 0.56\\
\hspace{1em}Diastolic blood pressure & 70 & 70 & -0.01 & 0.99\\
\hspace{1em}Respiratory rate & 16 & 17 & -0.11 & 0.78\\
\hspace{1em}Pre-treatment heart rate & 66 & 73 & -0.80 & 0.07\\
\midrule
\addlinespace[0.3em]
\multicolumn{5}{l}{\textbf{Outcome}}\\
\hspace{1em}Post-treatment heart rate & 78 & 87 & -0.92 & 0.02\\
\bottomrule
\end{tabular}
\begin{tablenotes}
\item Note: For continuous variables, the standardized difference is defined as $(\overline x_t-\overline x_c)/\sqrt{(s_t^2+s_c^2)/2}$, where $\overline x_t$ and $s_t^2$ denote the sample mean and the sample variance in the treatment group, respectively, and $\overline x_c$ and $s_c^2$ denote the counterpart in the control group, respectively. For binary variables, the standardized difference is defined as $(\widehat p_t-\widehat p_c)/\sqrt{(\widehat p_t(1-\widehat p_t)+\widehat p_c(1-\widehat p_c))/2}$, where $\widehat p_t$ and $\widehat p_c$ denote the sample mean in the treatment and control groups, respectively. $p$-values are computed by $t$-tests.
\end{tablenotes}
\end{threeparttable}
\end{table}

\begin{table}[H]

\caption{\label{tab:rd_main}Statistical and computational performance of different methods in two clinical trial datasets.}
\centering
\begin{threeparttable}
\begin{tabular}[t]{lrrrrrrrrrrrr}
\toprule
\multicolumn{1}{c}{ } & \multicolumn{4}{c}{Design ($\times 10^{-2}$)} & \multicolumn{5}{c}{Inference ($\times 10^{-2}$)} & \multicolumn{3}{c}{Run time (sec.)} \\
\cmidrule(l{3pt}r{3pt}){2-5} \cmidrule(l{3pt}r{3pt}){6-10} \cmidrule(l{3pt}r{3pt}){11-13}
\multicolumn{1}{c}{Method} & \multicolumn{1}{c}{$M$} & \multicolumn{1}{c}{$E_n$} & \multicolumn{1}{c}{$D_n$} & \multicolumn{1}{c}{$L_n$} & \multicolumn{1}{c}{Bias} & \multicolumn{1}{c}{SD} & \multicolumn{1}{c}{Size} & \multicolumn{1}{c}{CP} & \multicolumn{1}{c}{Len.} & \multicolumn{1}{c}{Sample} & \multicolumn{1}{c}{Ex.} & \multicolumn{1}{c}{Bi.}\\
\midrule
\addlinespace[0.3em]
\multicolumn{13}{c}{\textbf{The phase 1 clinical trial}}\\
\hspace{1em}CR & 700 & 100 & 3 & 137 & 7.35 & 414.5 & 5.1 & 94.9 & 1704 & 0.1 & 0.01 & 0.62\\
\hspace{1em}GPS & 43 & 99 & 12 & 222 & 9.45 & 338.5 & 5.4 & 94.6 & 1368 & 17.2 & 0.01 & 0.61\\
\hspace{1em}RR & 63 & 99 & 12 & 231 & 7.13 & 331.2 & 4.3 & 95.7 & 1377 & 115.8 & 0.01 & 0.61\\
\hspace{1em}PSRR & 63 & 99 & 12 & 222 & 9.65 & 341.9 & 6.0 & 94.0 & 1371 & 5.6 & 0.01 & 0.62\\
\midrule
\addlinespace[0.3em]
\multicolumn{13}{c}{\textbf{The internet-delivered clinical trial}}\\
\hspace{1em}SeqCR & 494 & 100 & 3 & 293 & 0.06 & 4.6 & 4.8 & 95.2 & 18 & 1.7 & 0.04 & 0.94\\
\hspace{1em}SeqRR & 1 & 100 & 4 & 351 & 0.03 & 4.0 & 3.2 & 96.8 & 16 & 86.7 & 0.04 & 1.01\\
\hspace{1em}SeqPSRR & 1 & 100 & 4 & 349 & 0.15 & 4.0 & 3.2 & 96.8 & 16 & 33.4 & 0.04 & 0.95\\
\bottomrule
\end{tabular}
\begin{tablenotes}
\item Note: $M$, mean Mahalanobis distance; $E_n$, $D_n$, and $L_n$, randomness metrics; Bias, absolute bias; SD, standard deviation; CP, coverage probability; Len., mean interval length; Sample, sampling $10^3$ acceptable assignments; Ex./Bi., determining interval endpoints by the exact approach/bisection method.
\end{tablenotes}
\end{threeparttable}
\end{table}

\end{document}